\documentclass[10pt]{article}
\usepackage{multirow,amsmath,amssymb}
\usepackage{graphicx}  
\def\ind{{\mathchoice {\rm 1\mskip-4mu l} {\rm 1\mskip-4mu l}
{\rm 1\mskip-4.5mu l} {\rm 1\mskip-5mu l}}}
\newtheorem{definition} {Definition} 
\newtheorem{proposition} {Proposition} 
\newtheorem{example} {Example} 
\usepackage{tikz}
\usepackage{pgfplots}
\usepackage{smartdiagram}
\begin{document}

\title{How Much Does Users' Psychology  Matter in  \\ Engineering  Mean-Field-Type Games}

\author{Giulia Rossi, Alain Tcheukam and Hamidou Tembine \thanks{Part of this work appeared in \cite{t1}.}
\thanks{The authors are with Learning \& Game Theory Laboratory, New York University Abu Dhabi, 
        {\tt\small tembine@nyu.edu}}%
}

\maketitle

\begin{abstract}

 Until now mean-field-type game theory was not  focused on cognitively-plausible models of choices in humans, animals, machines, robots, software-defined and mobile devices  strategic interactions. 
This work presents   some effects of users' psychology in mean-field-type games.  In addition  to the traditional ``material" payoff modelling, psychological  patterns are introduced in order to better capture and understand  behaviors that are observed in engineering practice or in experimental settings.   
The psychological payoff value  depends  upon choices, mean-field states, mean-field actions, empathy and  beliefs.  It is shown that  the affective empathy enforces mean-field equilibrium payoff equity and improves fairness between the players. It establishes equilibrium systems  for such  interactive decision-making problems. Basic empathy concepts  are illustrated in  several important problems in engineering including resource sharing, packet collision minimization, energy markets, and forwarding in Device-to-Device communications.
The work conducts also an experiment with 47 people who have to decide whether to cooperate or not. The basic  Interpersonal Reactivity Index of empathy metrics were used to measure the empathy distribution of each participant. Android app called Empathizer is developed to analyze systematically the data obtained from the participants. The experimental results reveal that the dominated strategies of the classical game theory are not dominated any more when users' psychology is involved, and a significant level of cooperation is observed among the users who are positively partially empathetic.

\end{abstract}

\newpage
\tableofcontents

\newpage

\section{Introduction}
Until now, mean-field-type game theory was not focused on cognitively-plausible models of choices in  humans, animals, machines, robots, software-defined and mobile devices  strategic interactions. This paper studies  behavioral and psychological games of mean-field type. Psychological  games seems to explain behaviors that are better captured in experiments or in practice than classical game-theoretic equilibrium analysis. It takes in consideration  psychological patterns of the decision-makers in addition to the traditional ``material" payoff modelling. The payoff value  depends  upon choice consequences, mean-field states, mean-field actions and on beliefs about what will happen. The psychological game theory framework can link cognition, emotion, and express emotions, guilt, empathy, altruism, spitefulness (maliciousness) of the decision-makers. It also include belief-dependent and other-regarding preferences in the motivations. 

One motivating example of psychological game theory is  trying to understand how her users and consumers  will perceive a product or are thinking about a product  in web online shop  and will engage in empathy in the interaction. There are several definitions of empathy in the literature (see \cite{prest}). Cognitive empathy of a player, sometimes also called perspective taking, is the ability to identify the felling and emotions of other players.  Perspective taking empathy is considered as the experience of understanding another player's state and actions from their perspective or mutual perspective via several channels. A decision-maker can place herself in the shoes of the others  and feel what they are feeling. This is a
 particularly useful concept in the context of psychological game theory. Indeed, it helps to anticipate, compute and to react to the behavior of the others thanks to empathy. 
Note that, empathy is different than sympathy which is  the ability to select appropriate emotional responses for the apparent emotional states of others.
In other words, sympathy is not about feeling the same thing that somebody else is feeling, but an appropriate emotion to complement theirs. Another notion is compassion 
which heuristically is to treat others as you would like to be treated. It consists in selecting the appropriate action in response to the apparent emotional states of another. This active version of empathy may result in partial altruism in the preferences formation of the players.  In game theory, the strategy and the resulting actions play key roles in the outcomes. A player may use empathy in different ways. Examples include empathy-selfishness, empathy-altruism and empathy-spitefulness.  In this work, we examine basic empathy subscales: perspective taking (PT), empathy concern (EC), fantasy scale (FS) and personal distress (PD) that will be evaluated through the Interpersonal Reactivity Index (IRI, \cite{davis1}).

\subsection{Overview}
We overview some prior works on empathy in game theory.
The motivation of decision makers who care for various emotions, intentions-based reciprocity, or the opinions of others may depend directly on beliefs (about choices, states, behaviors, or information).  
The study in \cite{prest} tries to explain how we can understand what someone else feels when he or she experiences simple emotions.  Some psychological factors are considered in a game theoretic context \cite{psy1,psy2,psy3,psy5,psy6,ber1}.  The work in \cite{nref1} investigates the neural basis of  complex decision making using a game theory.   The authors of
\cite{nref2} study  neuroeconomics approach to decision-making by combining  game theory with  psychological and neuroscientific methods.
The work in \cite{ber2} underlined how empathy leads to fairness and \cite{ber3} studied the correlation between empathy, anticipated guilt and prosocial behaviour; in his study he found out that empathy affects prosocial behaviour in a more complex way than the one represented by the classical model of social choices.
The authors in \cite{psy2}  propose and synthesize a large body of experimental and theoretical analysis on multi-agent interactions, in psychology 
as well as economics. The work in
\cite{nref3}  presents a game theoretic approach to empathy, reporting on current knowledge of the evolutionary, social, developmental, cognitive, and 
neurobiological aspects of empathy and linking this capacity to human communication, including in clinical practice and medical education.
Under mild conditions, \cite{har} shows  that such empathetic preferences requiring us to see things from another's  point of view can be summarized by 
empathic payoff function. The idea of empathic preferences was followed in \cite{nref4}  where the importance of empathic payoff function is illustrated. 
The work in  \cite{nref7} presents an overview on empathy and mind reading in
some detail and have pointed out  other-regarding preferences in game theory.
In the classical approach, they are taken to be, in the
'worst' case, a purely selfish or, in the best case, self-
regarding individuals. In the psychological payoffs, players
may care about the betterment of others as well as
themselves but their  position, state or consumption is simply
another argument of their own preferences. Other-regarding preferences are sometimes presented not
simply as being concerned with the payoff  of specific others, but may incorporate more general concerns,
such as equity, risk and fairness. The problem of comparing the payoff of different players is vital for the question of fairness and equity \cite{nref5}.
 The work in \cite{ber4} proposes an evolutionary game-theoretic approach to study the evolutionary effect of empathy on cooperative games. 
The origin, the source and learnability of   empathy  remains open. The evolution of empathy and its connection  to reciprocal altruism are discussed  in \cite{nref6}. 
Building  on dynamic interactive epistemology, \cite{psy4}  proposes a more 
general framework that includes  higher-order beliefs, beliefs of others, and plans of action may influence motivation, 
dynamic psychological patterns (such as sequential reciprocity, psychological forward induction, and regret). These preliminary studies enrich classical game theory by empirical knowledge and makes it  significantly closer to what is needed for real-world applications. Thousands of theoretical papers have been published about the prisoner's dilemma game and more than 30  experiments about  have been conducted in the literature,  only few of them are dedicated to the emergence of cooperative behaviors in 
one-shot games \cite{batson01}.  The works in \cite{refsmc1,refsmc2,refsmc3} consider decision-making problems. However, the effect of users' psychology on its decision is not examined in these previous works. None of these previous works considered a mean-field-type game setup. Finally, the range of applications covered by these papers is limited compared to the current work.

The goal of this paper is to examine the effect of empathy on players' behavior and outcomes in mean-field-type game theory. The psychology of the players is analyzed in several engineering applications.

\subsection{Contribution}
Our contribution can be summarized as follows.
This paper  illustrates how some insights from the psychology literature on empathy can be incorporated into a mean-field-type payoff function, and demonstrate the potential interaction of beliefs, strategies, mean-field through the channel of empathy. It establishes  mean-field equilibrium systems with psychological payoffs. 

\begin{itemize}\item Empathy as perspective taking may  induce a partial altruism. To a partially altruistic player we would be considering a payoff in the form
${r}_{j}^{\lambda}=r_j+\sum_{j'\neq j} \lambda_{jj'}r_{j'},
$ where $\lambda_{jj'}\geq 0.$

It is shown that empathy-altruism promotes fairness in terms of mean-field equilibrium payoffs in  wide range of mean-field-type games.    We illustrate empathy concepts in several
 important engineering problems. \begin{itemize}\item  We  provide an experimental evidence that the degree of empathy  can shift the decisional balance in a  one-shot  forwarding dilemma game. To this end, we have conducted an experimental test with 47  people carrying mobile devices. We have developed a android app called empathizer to measure dimensional empathy of the participant (Fig.\ref{fig:empathizer}). It turns out that empathy induces some cooperative behaviors which consist  to forward the data of the other wireless nodes in a network. Applied to D2D communications and WiFi Direct technology, this experiment helps us to estimate the proportion of users who are potentially interesting in enabling their platform to the others. The same method can be useful in other contexts such as mobile crowd sensing 
 which pertains
to the monitoring of large-scale phenomena that
cannot be easily measured by a single individual user.
For example, intelligent transportation systems
may require traffic congestion monitoring and air
pollution level monitoring. These phenomena can
be measured accurately only when many individuals
provide speed and air quality information from
their daily commutes, which are then aggregated spatio-temporally to determine congestion and pollution levels in smart cities.
It is thus important to estimate the number of potential participants who decide their level of participation to the crowdsensing when these users are carrying power-hungry devices to serve the cloud data. 
\begin{figure}[htb]
\centering
\includegraphics[scale=0.5]{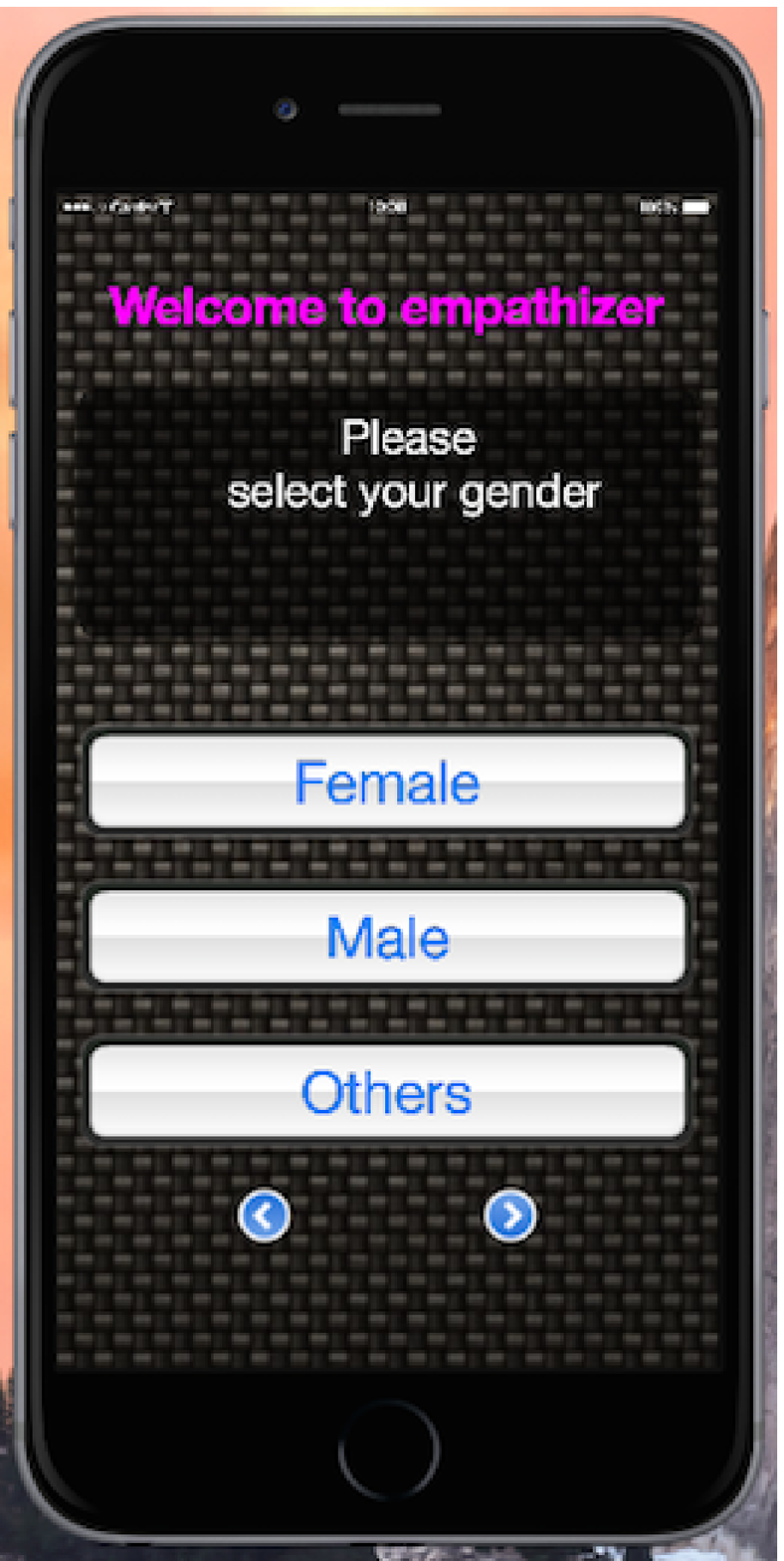}\  \includegraphics[scale=0.5]{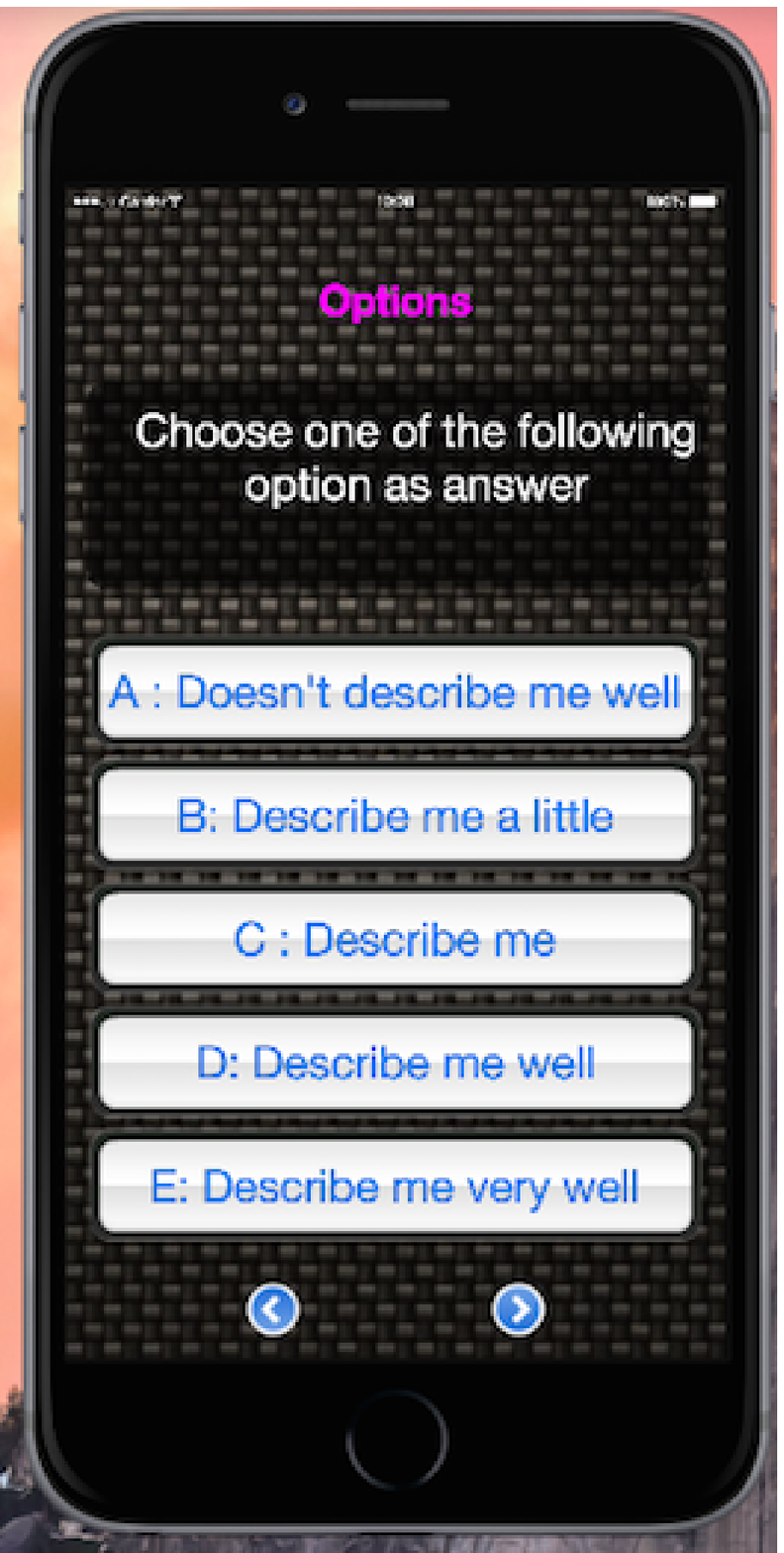}\caption{Empathizer app: Sample welcome screen on android platform for measuring empathy at NYUAD L\&G Lab.}
\label{fig:empathizer}
\end{figure}

   \item  Empathy-altruism provides a better explanation of resource sharing outcomes. \item The empathy-altruism of the users may help in reducing
packets collision in wireless medium access channel and hence reducing congestion. \item Empathy-altruism reduces energy consumption during  peak hours. \item 
 Empathy-altruism of prosumers improves their equilibrium revenues. 
 \end{itemize}

\item 
Can empathy at times be harmful?  

We do not restrict ourselves to the positive part of empathy. Empathy may have a 'dark' or at least costly side specially when the environment is strategic and interactive as 
it is the case in games. 

Can empathy be bad for the self?   Empathy can be used, for example, by a other player attacker to identify the weak nodes in the network.  

Can empathy be bad for others? Empathetic users may use their ability to destroy the opponents.   In strategic interaction between people,
empathy may be used to derive  antipathetic response (distress at seeing others' pleasure, or pleasure at seeing others' distress).

 We illustrate this in a context of auction in prosumer (consumer-producer) markets.
A prosumer who is bidder might be losing the auction due highly competitive prices. Yet she
 participates in the auction because she wants to minimize the negative payoff on losing by making
her competitor, who would win the auction, gets low reward by selling its energy at almost zero price or negative price and hence the other get a high  price for the win. This negative dependence of payoff
on others' surplus is referred to as spiteful behavior. We associate a certain spitefulness coefficient $-\alpha_{jj'}\leq 0$ to the bidder $j.$
A spiteful player $j$ maximizes the weighted difference of her
own profit $r_j$ and his competitors' profits $r_{j'}$ for all $j'\neq j.$
The payoff of a spiteful (antipathetic) player is
\begin{eqnarray}
{r}_{j}^{\alpha}:=\alpha_j r_j-\sum_{j'\neq j}\alpha_{jj'}r_{j'}
\end{eqnarray}
Obviously, setting $\alpha_{jj'}=0, \ j\neq j$ and $\alpha_j=1$
 yields a selfishness
 (whose payoff equals his exact profit) whereas $\alpha_j=0,\ \alpha_{jj'} = 1$ defines
a  malicious player (jammer) whose only goal is to minimize
the global profit of other players. 
For $\alpha_j\neq 0,$ we can scale the payoff by $\frac{1}{\alpha_j},$ to get
$\alpha_j\left[r_j-\sum_{j'\neq j}\frac{\alpha_{jj'}}{\alpha_j}r_{j'}\right],$ which is equivalent to focusing on
$$
{r}_{j}^{\lambda}=r_j-\sum_{j'\neq j}\lambda_{jj'}r_{j'}
$$
where $\lambda_{jj'}:=\frac{\alpha_{jj'}}{\alpha_j}.$ This class of games captures a very extreme scenario in which everyone dislikes all the others. It is shown that  the empathy-spitefulness of prosumers decreases the optimal bidding  price of the winners.  This means that the spitefulness of the prosumers may benefit to the consumers.

\end{itemize}

 \subsection{Structure}
 
 The rest of the paper is organized as follows. 
 In Section \ref{sec:motivation} we provide motivating examples illustrating how empathy-altruism improves fairness. A generic mean-field-type game is presented  and analyzed in Section \ref{sec:model}. Section \ref{sec:example} illustrates empathy in variance reduction problems. Section \ref{sec:experiment} presents an experimental setup for participation in data forwarding in D2D communications. 
 Section \ref{sec:conclusion}  concludes the paper.

 Notations used in the text are available in Table \ref{tablenotatt}.
 \begin{table}
 \begin{tabular}{|c|c|}
\hline
Meaning &Notation\\
\hline
Horizon & $\{0, 1,\ldots, T-1\}$\\
Number of players & $n$\\ 
Time index & $t$\\
Noise & $\eta_t$\\
State & $s_t$\\
State mean-field& $m_t^s$\\
Action profile & $a_t$\\
Action mean-field & $m^a_t$\\
Instant material payoff of $i $ & $ r_{it}(s_t, m_t^s, m_t^a, a_t)$\\
Terminal material payoff of  $i$ & $g_{iT}(s_T, m_T^s)$\\
Instant empathic payoff of $i $ & $r_{it}^{\lambda}(s_t, m_t^s, m_t^a, a_t)$\\
Terminal empathic payoff of  $i$  & $g_{iT}^{\lambda}(s_T, m_T^s)$\\
Degree of empathy of  $i$ &  $\lambda_i\in \mathbb{R}^{n-1}$ \\
\hline
\end{tabular} \label{tablenotatt}  \caption{Notations }
\end{table}

\section{Empathy in game theory} \label{sec:motivation}

\subsection{Empathetic preferences}
We consider the local empathetic preference, the outcome of agent $i$ for action $a_i$ combines her intrinsic preference for $a_i$ with the intrinsic preference of $i$'s neighbors, $\mathcal{N}_i$, where the weight given to the preference of any neighbor $j\in \mathcal{N}_i$ depends on the strength of the relationship between $i$ and $j.$ A basic setup and for illustration purpose this can be captured with a number $\lambda_{ij}$ but it could be a general map with beliefs.

\subsection{Empathic payoffs}
\subsubsection{Self-regarding payoffs}
By a self-regarding player we refer to a player  in a game  who maximizes his own payoff $r_i.$ A self-regarding player $i$ 
 thus cares about the behavior  and payoffs that impact her own payoff $r_i.$ 
\subsubsection{Other-regarding payoffs}
 An other-regarding player $i$ considers not only her own payoff $r_i$ but also some of her network members' payoffs $(r_j)_{j\in \mathcal{N}_i}.$ Then, the player will include these in her preferences and create an empathic payoff. She is still acting to maximize her new empathic payoff.

\subsubsection{Reciprocity payoffs} 
There are many interactive decision-making situations where both positive and negative reciprocal behaviors are observed.  A  user carrying a wireless device may favorably accept to forward the data of another temporary device (a new joiner at a public place,  conference or airport), and that device reciprocates the favor although it is unlikely that they will ever meet again. In order to capture such a phenomenon in the preferences, the kindness between  players and  higher order reciprocity terms \cite{rec1,rec2} will be introduced below.

\subsubsection{Evolution of empathy}
The question of whether there is a fixed distribution
of degrees of other-regarding behaviour in the network is important.  We  investigate the effect of dynamic empathy on the payoffs in mean-field-type games. The basic experiment reported below reveals that there  is a distribution of empathy  across the population (see Figure \ref{fig:repartition} and  Table \ref{table:IRIdistr1}) and it is context-specific.

\subsection{Motivating Examples}  \label{sec:motivation2}
In this section we discuss the interplay of self-regarding and other-regarding behavior through  motivating examples. 
\subsubsection{Empathy Explains Better the Cake Splitting Behavior}
As inspired by the Ultimatum Game, we consider two sisters who are asked to split/share a cake. The first sister, the proposer, makes an offer of how to split the cake. The second sister, the responder, either accepts the offer, in which case the cake is split as agreed, or rejects it, in which case neither sister receives anything and a restart the process. If we model only with material payoff, and the horizon is $1$ then a good strategy  for the proposer is to offer the smallest possible positive share of cake and for the responder sister to accept it. However the material payoff is not what is widely observed in engineering practice. Why? One possible explanation may come from psychology using the empathy of the sisters for each other. The sisters do not behave this way, however, and instead tend to offer nearly 50\% of the cake and to reject offers below 20\%.  Empathy means that individuals make offers which they themselves would be prepared to accept. If only below 20\% of the cake were proposed to yourself you would not accept so you will not propose an offer below 20\%.
Following that idea we will see that empathy can lead to the evolution of fairness in the interaction.

\subsubsection{Empathy-Altruism Reduces Collisions in Wireless Channels}
Consider $n$ wireless devices sharing a common medium channel using Aloha-like protocol.
If two or more users transmit simultaneously, there is a collision and the packets are lost. If only one user transmits at a time slot with transmission power $\bar{p}>0$ then the transmission is successful if the received signal is good: the signal-to-noise ratio (SNR) is above a certain threshold $\beta_i.$ The success condition is $ \ind_{SNR_i\geq \beta_i}$ which is a random variable which is equal to $1$ if  $SNR_i\geq \beta_i$ and $0$ otherwise.
\begin{figure}[htb]
\begin{tabular}{cc|c|c|c|c|l}
\cline{3-4}
& & \multicolumn{2}{ c| }{Player I} \\ \cline{3-4}
& & Transmit & Wait  \\ \cline{1-4}
\multicolumn{1}{ |c  }{\multirow{2}{*}{Player II} } &
\multicolumn{1}{ |c| }{Transmit} & (0,0) & $(\ind_{SNR_1\geq \beta_1},0) $      \\ \cline{2-4}
\multicolumn{1}{ |c  }{}                        &
\multicolumn{1}{ |c| }{Wait} & $(0,\ind_{SNR_2\geq \beta_2})$ & (0,0)    \\
 \cline{1-4}
\end{tabular}
\label{fig:rmg1}
\caption{Random payoff matrix of wireless collision channel game with self-regarding players. ``T" is for Transmit and ``W" for Wait. }
\end{figure}
This is an interactive decision-making framework for channel access point in wireless networks where the outcome is influenced not only by the decisions of the users but also by a random variable representing the channel state. This belongs to the class of random matrix games (RMGs, \cite{rmg}) because the SNR appears in the entries of the payoff-matrix as the SNR depends on the channel  state which is random process.

Considering the expected material payoff, it is not difficult to observe that $ (\delta_{T},y\delta_{T}+(1-y)\delta_{W}) $ is an equilibrium for any $y\in [0,1].$ In particular the pure action profile $(T,T)$ is an equilibrium. Thus the payoff gap between the payoffs in equilibrium is  $$IN(0)=\max\{ \mathbb{P}(SNR_i\geq \beta_i), \ i\in \{1,2\}\}.$$

We denote by $\lambda_i$ the degree of empathy-altruism of user $i.$ Then, the empathic-altruism payoff matrix  is given by
\begin{figure}[htb]

\begin{tabular}{cc|c|c|c|c|l}
\cline{3-4}
& & \multicolumn{2}{ c| }{Player I} \\ \cline{3-4}
& & T& W  \\ \cline{1-4}
\multicolumn{1}{ |c  }{\multirow{2}{*}{Player II} } &
\multicolumn{1}{ |c| }{T} & (0,0) & $ (a_{12},b_{12})  $      \\ \cline{2-4}
\multicolumn{1}{ |c  }{}                        &
\multicolumn{1}{ |c| }{W} & $(a_{21}, b_{21})  $ & (0,0)    \\
 \cline{1-4}
\end{tabular}

\label{fig:rmg}
\caption{Random payoff matrix of wireless collision channel game with partial empathy-altruism. }
\end{figure}

where $$ (a_{12},b_{12})=(\ind_{SNR_1\geq \beta_1},\lambda_2 \ind_{SNR_1\geq \beta_1}),$$
$$(a_{21}, b_{21})=(\lambda_1\ind_{SNR_2\geq \beta_2},\ind_{SNR_2\geq \beta_2}).$$

\begin{figure}[htb]
\includegraphics[scale=0.25]{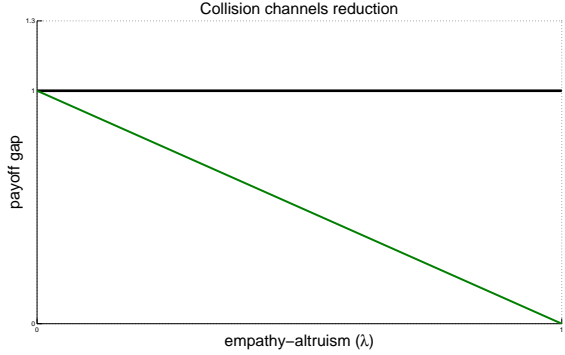}\\ \caption{Impact of empathy-altruism  on  collision channel reduction. As the altruism level of the users increases the collision ratio decreases. }
\label{fig:collision}
\end{figure}

With this expected empathic payoff, the profile $ (T,{W}) $ is an equilibrium and  the pure action profile $(T,T)$ is no longer an  equilibrium if  $\max\{ \mathbb{P}(SNR_i\geq \beta_i), \ i\in \{1,2\}\}>0.$ Thus, the equilibrium payoff gap between the users  is  $$IN(\lambda)=\max\{ (1-\lambda_j)\mathbb{P}(SNR_i\geq \beta_i), \ i,j\in \{1,2\}\}$$ which is smaller than $IN(0)$ for any $\lambda\in (0,1).$ This says that the concept of empathy-altruism helps to reduce collisions in wireless medium access control (see Figure \ref{fig:collision}).

\subsubsection{Empathy-altruism reduces peak hours energy consumption}

Consider  $n$  consumers interacting in an energy market.  The electricity price is a function of the aggregated supply $S$ (of the producers) , the aggregated demand  $D$ (of the consumers) and on mismatch between supply and demand.  The payoff $r_i$ of  consumer $i$ depends on the amount of energy consumed $d_i$, on  his/her own degree of satisfaction $w_i(d_i)$ (a typical satisfaction function would be  $w_i(d_i) := 1- e^{-d_i}$) and on the electricity price $p.$ Let $ r_i =  w_i(d_i) - p(D,S)d_i$ be the material payoff. 
The interior equilibrium (if any)  when users are empathic-selfish satisfies $w'_i -p'_{d_i}d_i=p$ and the solution is denoted by $d_i^*(0).$

\begin{figure}[htb]
\includegraphics[scale=0.3]{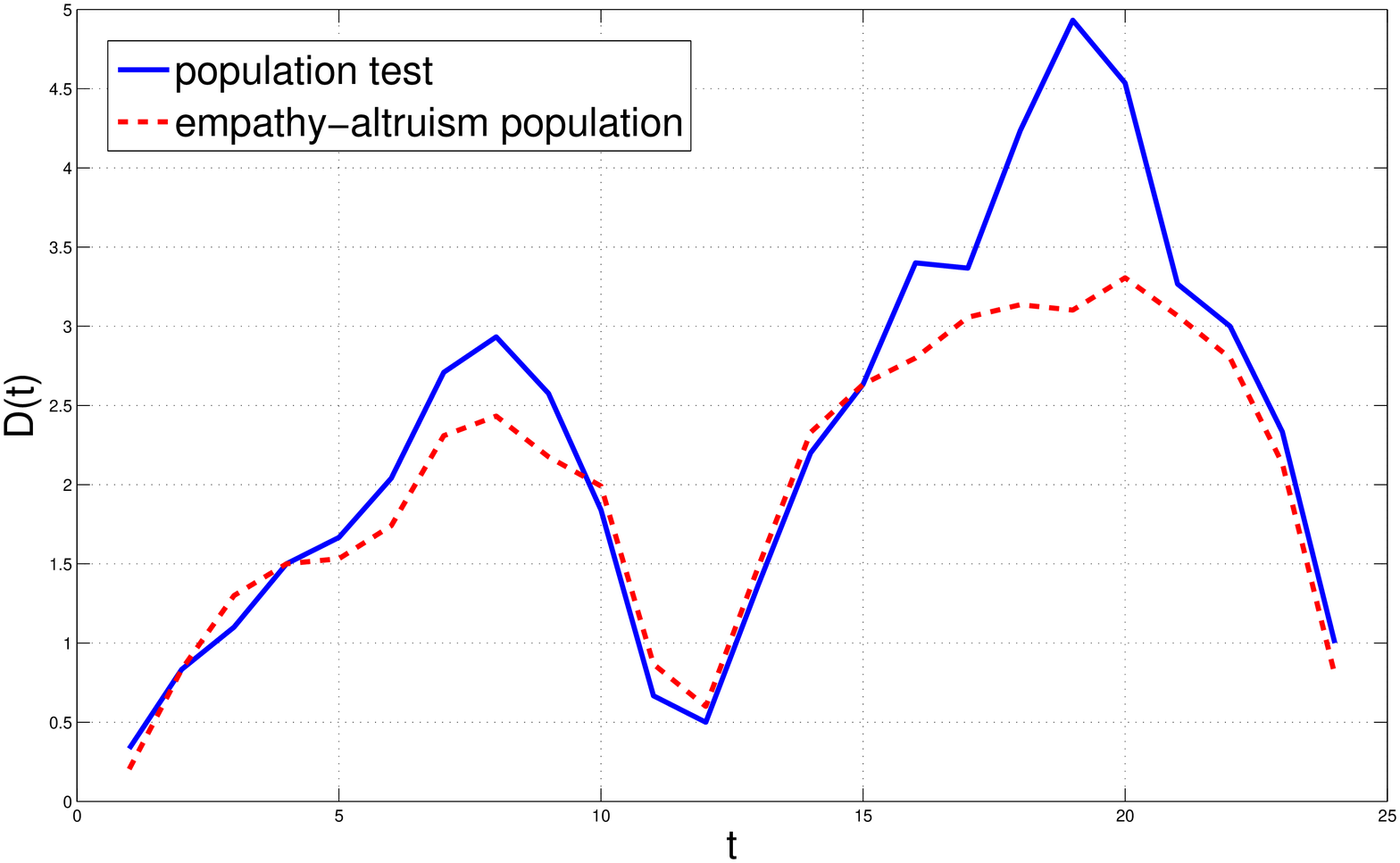}\\ \caption{Impact of empathy-altruism  on  peak energy demand reduction}
\label{fig:energy}
\end{figure}

Let $\lambda \in (0,1)$ be a parameter modeling the degree of empathy-altruism of a consumer in the power network.  We denote by  $\tilde{r}_i = r_i +  \lambda r_j$ the empathic payoff of prosumer $i$ given his/her empathy-altruism toward consumers $j,\  j \neq i.$  The interior equilibrium   when users are empathic-altruistic satisfies $$w'_i (d_i)-p'_{d_i}d_i=p-\lambda d_j,$$ that solution is denoted by $d_i^*(\lambda).$ Since $p-\lambda d_j\leq p$ and the function $d_i \mapsto  w'_i(d_i) -p'_{d_i}d_i$ is non-decreasing, and a one-to-one mapping within its range, it turns out $$ d_i^*(\lambda)\leq d_i^*(0), \forall \lambda\in (0,1).$$ Summing over the consumers, we obtain
$$  D^*(\lambda):=\sum_{i}d_i^*(\lambda)\leq D^*(0)=\sum_{i}d_i^*(0),$$
which means that empathy-altruism reduces the energy consumption. In particular it reduces the global peak demand during peak hours.  Figure \ref{fig:energy} represents the total demand curve $D(t)$ for 1 day. One can observe two important peaks which 
are significantly reduced when players are empathic. The question of how to incentivize users' to be more empathic is an interesting direction that we leave for future research.

\subsubsection{ Empathy-spitefulness of prosumers decreases the optimal bidding  price}
Each prosumer has a unit production cost and quantity $q.$
The production $c_j$ is a random variable with support in $[0,\bar{c}]$ and with cumulative function equals to $F(.).$ 
Each prosumer knows its own production cost, its spitefulness coefficient, its bid but not the production cost of the other bidders. Each bidder knows the cumulative distribution of the others.

The game is played as follows. Each prosumer bids a (unit) price $p_j$ with the production quantity $q_j>0.$ The expected material payoff is
$$r_{j}=q_j (p_j-c_j)\ind_{ ( p_j < \min_{j'\neq j}p_{j'}(c_{j'}) )}  -e_j,$$  where $e_j$ is the entry cost to the energy market.
The empathic-spiteful payoff of prosumer  $j$ is 
$r_{j}^{\lambda}=r_{j}-\lambda r_{i}$ with $\lambda>0.$

We  are interested in structural results of equilibrium strategy under spiteful coefficient. Let $X$ be a random variable drawn from the interval $[0, \bar{c}]$ with 
cumulative distribution function $I.$ Then, the conditional expectation of $X$ given that $X$ is greater than $c$ is given by
$$ \mathbb{E} (X \ | X>c)= \int_c^{\bar{c}} x \frac{I'(x)}{1-I(c)} \ dx.$$

The optimal bidding (price) strategy of the prosumer is 
$$
p^*(c,F,\lambda)= \mathbb{E} (X_{\lambda} \ | X_{\lambda}>c),
$$ where $X_{\lambda}$ is a random variable with cumulative function $$I_{\lambda}(c)=P(X_{\lambda}< c)=1-(1-F(c))^{1+\lambda}.$$

It can easily be checked that $I_{\lambda}(c)$ is indeed a valid cumulative distribution function
$I_{\lambda}(0) = 0, I_{\lambda}(1) = 1,$ and $I_{\lambda}$ is non-decreasing and differentiable. Note that the optimal bidding price of a winner is  likely to be above $c$ so that the prosumer gets some benefits in selling electricity to the market. The optimal bidding price $p^*(c,F,\lambda)$  decreases as  the spitefulness parameter $\lambda$ increases.
 \subsubsection{ Empathy-Altruism of Prosumers Improves the revenue of the Prosumers}
 We now examine the effect of  Empathy-Altruism in the revenue of the prosumers.
 From the above analysis, the altruism strategy is obtained by changing the sign of $\lambda.$ 
 The optimal bidding price $p^*(c,F,-\lambda)$  for partially altruistic  prosumers increases as $\lambda$ increases. This will help the prosumers to save more:  the benefit $p^*-c$ increases with $\lambda\in (0,1).$

In the next section we present a class of mean-field-type games \cite{temftg,temftg2,temftg3,temftg4} and explain how the above preliminary results in Subsections (1)-(5) on the psychology of the players can be extended to this context.
\section{Psychological mean-field-type games} \label{sec:model}

\begin{definition}[Mean-Field-Type Game] \label{defi1}
A mean-field-type game is a  game in which the instantaneous payoffs and/or the state dynamics coefficient functions involve  not only the state and the action profile but also the joint distributions of state-action pairs (or its marginal distributions, i.e., the distributions of states  or the distribution of actions). A typical example of payoff  function of player  $j$ has the following structure: $$r_j:\ \mathcal{S}\times A\times \mathbb{P}( \mathcal{S}\times A) \rightarrow \mathbb{R},$$ with $r_j(s,a, D_{(s,a)})$ where $(s,a)$ is the state-action profile of the players and $D_{(s,a)}$ is the distribution of the state-action pair $(s,a),$  $\mathcal{S}$ is the state space  and $A$ is the action profile space of all players. 
\end{definition}

From Definition \ref{defi1}, a mean-field-type  game can be static or dynamic in time.
In mean-field-type games, the number of players is arbitrary: it can be finite or infinite \cite{te1,te2,te3}.
The indistinguishability property (invariance in law by permutation of index of the players) is not assumed. A single player may have a non-negligible impact of the mean-field. This last property makes a strong difference between ``mean-field games" and ``mean-field-type games". 

One may think that ``mean-field-type games" is  a small and particular class of games. However, this class includes the classical games in strategic form because any payoff function $r_j(s, a)$ can be
 written as $r_j(s,a, D)$ where $D_{(s,a)}$ is the distribution of the state-action pair $(s,a).$   Thus, the form $r_j(s,a, D)$ is more general and includes non-von Neumann payoff functions.
  \begin{example}[Mean-variance payoff] The payoff function of agent $i$ is $E[r_i(s,a)]-\lambda \sqrt{var[r_i(x,a)]}, \lambda\in \mathbb{R}$ which can be written as a function of $r_i(s,a,D_{(s,a)}).$  For any number of interacting players, the term $D_{s_i,a_i)}$ plays a non-negligible role in the standard deviation $\sqrt{var[r_i(s,a)]}.$ Therefore, the impact of agent $i$ in the individual mean-field term   $D_{(s_i,a_i)}$ cannot be neglected.
\end{example}

\begin{example}[Aggregative games] 
The payoff function of each player depends on its own action and an aggregative term of the other actions. Example of payoff functions include  $r_i(a_i, \sum_{j\neq i} a^{\alpha}_j), \ \alpha>0$ and 
 $r_i(s_ia_i, \sum_{j\neq i} s_ja_j).$ 
\end{example}

 In the non-atomic setting, the influence of an individual state $s_j$ and individual action $a_j$ of any user $j$ will have a negligible impact on mean-field term  $D_{(s,a)}.$ In  that case, one gets to the so-called mean-field game.

\begin{example}[Population games] Consider a large population of agents. Each agent has a certain state/type $s \in \mathcal{S}$ and can choose a control action $a\in \mathcal{A}(s).$ Let $m$ the proportion of type-action of the population.
The payoff of the agent with type/state $s,$ control action $a$ when the population profile $m$ is $r(s, a,m).$  
{\it Global games} with continuum of players is based on the Bayesian games and uses  the proportion of actions (mean-field of actions).
\end{example}

 In the case where  both non-atomic and atomic terms are involved in the payoff, one can write the payoff  function as $r_j(s,a, D, \hat{D})$ where $\hat{D}$ is the population state-action measure. User $j$ may  influence $D_j$ (distribution of its own state-action pairs) but  its influence on $ \hat{D}$  may be limited.

\subsection{Psychological payoffs}

\subsubsection*{ Empathic payoff}
The instant empathic payoff of $i$ is $$ r_{i}^{\lambda}(s, m^s, m^a, a):= r_{i}+\sum_{j\in \mathcal{N}_i \backslash \{i\}}\lambda_{ij}r_{j}.$$

\begin{figure}[htb]
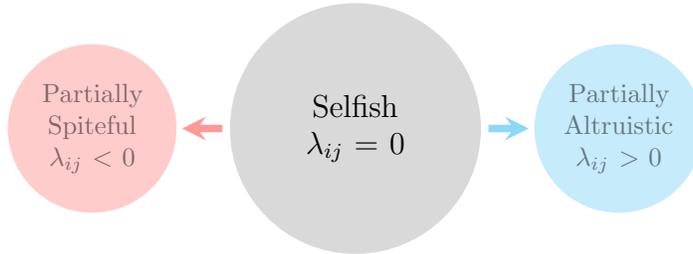

\smartdiagram[constellation diagram]{Selfish $\lambda_{ij}=0$,
  Partially Spiteful $\lambda_{ij}<0$, Partially Altruistic $\lambda_{ij}>0$}
  \label{fig:diagram}
  \caption{Behavior of $i$ towards $j$ for different sign values of $\lambda_{ij}.$ }
  \end{figure}

\begin{itemize} 
\item Selfishness:  If $\lambda_{ij}=0$ we say that $i$ is empathic-selfish towards $j.$ Player $i$ is self-regarding if $\lambda_{ij}=0$ for all $j\neq i.$
If all the $\lambda_{ij}$ are zeros for every $i,j$ then every player focuses on her own-payoff functions. 
\item Partially Altruistic: If $\lambda_{ij}\in (0,1)$ we say that $i$ is partially empathic-altruistic towards $j.$
If all the $\lambda_{ij}$ are positive  for every $i,j $ every player is considering the other players in its decision in a partially altruistic way.
\item Partially Spiteful/Malicious:  If $\lambda_{ij}< 0$ we say that $i$ is partially empathic-spiteful towards $j.$
If all the $\lambda_{ij}$ are negative  for every $i,j $ every player is considering the other players in her decision in a partially spiteful way.
\item Mixed altruism-spitefulness-neutrality:  The same player $i$ may have different empathetic behaviors towards her neighbors. If $\lambda_{ij}>0,$  $\lambda_{ik}<0$  and $\lambda_{il}=0$ for $j,k,l\in \mathcal{N}_i\backslash \{i\}$ then player $i$ is partially altruistic towards $j,$ and partially spiteful towards $k$ and neutral towards $l.$
\end{itemize}
\subsubsection*{ Reciprocity payoff}

Define $i$'s  kindness  to player $j$  as
$$\kappa_{ij}(a_i, (b_{ij})_{j\neq i})=r_j(a_i, (b_{ij})_{j\neq i})$$ $$-\frac{1}{2}\left[ \sup_{a'_i}r_j(a'_i, (b_{ij})_{j\neq i}) + \inf_{a'_i}r_j(a'_i, (b_{ij})_{j\neq i}) \right]$$
where $r_j(a_i, (b_{ij})_{j\neq i})$ the material payoff that player $i$ believes that player $j$ will receive. We say that $i$ is kind to $j$ if $\kappa_{ij}>0.$  $i$ is unkind to $j$ if  $\kappa_{ij}<0.$  $b_{ij}$ is $i$'s belief on player $j$'s strategy. 
 In order to define reciprocity, we introduce a second order reasoning.  Let $\tilde{b}_{ijk}$ is $k$'s belief about others. 
 
 The reciprocal perceived kindness of $j$ towards $i$ is
 $$\tilde{\kappa}_{iji}= r_i(b_{ij}, (\tilde{b}_{ijk})_{k\neq j})$$ $$-\frac{1}{2}\left[  \sup_{b'_{ij}}r_i(b'_{ij}, (\tilde{b}_{ijk})_{k\neq j}) + \inf_{b'_{ij}}r_i(b'_{ij}, (\tilde{b}_{ijk})_{k\neq j})\right]$$
 which is what $i$ believes that $j$ believes that $i$ will receive. If $\tilde{\kappa}_{iji}>0$ it means that $i$ perceives that $j$ is kind to him.
 
 The empathetic reciprocity payoff is
 $r_{i}^{\lambda}(a, b, \tilde{b})= r_i+ \sum_{j\in \mathcal{N}_i \backslash \{i\}} \lambda_{ij} \kappa_{ij}  \tilde{\kappa}_{iji},$
 where $\lambda_{ij}$ is $i$'s reciprocity sensitivity towards $j.$ If $\lambda_{ij}>0$ then $\kappa_{ij}. \tilde{\kappa}_{iji} >0$  (same sign) reflects mutual kindness or mutual unkindness.

An equilibrium requires a best response property and a consistency between these beliefs $ m^{a_k}=b_{jk}=\tilde{b}_{ijk}.$

\subsection{Basic Dynamic  Game Model}
Consider  a  dynamic mean-field-type game setup with the following data:
{\small $$\left\{ \begin{array}{cc}
\mbox{Time step: } & t\leq T\\
\mbox{Set of Players:} & \{1,\ldots, n\}\\
\mbox{Initial state : } &  s_{0} \sim m_0\\
\mbox{State dynamics: } &  s_{t+1} \sim q_{t+1}(. | \ s_t, m_t^s, m_t^a, a_t)\\
\mbox{Instant material payoff of}\ i : & r_{it}(s_t, m_t^s, m_t^a, a_t)\\
\mbox{Terminal material payoff of } i:  & g_{iT}(s_T, m_T^s)\\
\mbox{Instant psychological payoff of}\ i : & r_{it}^{\lambda}(s_t, m_t^s, m_t^a, a_t)\\
\mbox{Terminal psychological payoff of } i:  & g_{iT}^{\lambda}(s_T, m_T^s)\\
\mbox{Degree of empathy/reciprocity of } i: &  \lambda_i =(\lambda_{ij})_j \\
\end{array}
\right.
$$
}
where $T$ is the duration  of the interaction, $a_t=(a_{1t},\ldots,a_{nt})=: (a_{it},a_{-i,t}) $ represents a control-action profile of all players at time $t.$ $a_{i,t}\in A_i,$ the space of actions of  $i$ at time $t,$ $m^s_t$ is the distribution of state at time  $t$, $m^a_t$ is the distribution of actions at time $t.$

\begin{definition}[Behavioral pure strategy] A behavioral pure strategy of  player  $i$ at time $t$ is a mapping from the available information to the set of actions. The set of pure strategies of $i$ is denoted by $\mathcal{A}_i.$
\end{definition}

Player $i$'s cumulative empathic payoff is  $$R_i^{\lambda}(m_0^s,a)= E \sum_{t=0}^{T-1}r_{it}^{\lambda}(s_t, m_t^s, m_t^a, a_t) + g_{iT}^{\lambda}(s_T, m^s_T).$$

Next we define the response of a player to the others and the mean-field.
\begin{definition}[Best response]
A strategy $a_i$  of player $i$ is a best-response  to  $ (a_{-i}, m^{a_{-i}})$ if 
$$
R_i^{\lambda}(a)=\sup_{a'_i}  R_i^{\lambda}(a'_i, a_{-i}).
$$
The set of best response strategies of  player $i$ defines a best response correspondence $BR_i^{\lambda}.$
\end{definition}

The existence of a pure best-response strategy can be obtained in number of classes of games. When a pure best response strategy fails to exist, one can  use behavioral mixed strategies. Using weak compactness of the set of probabilities on ${A}_i,$ the existence of mixed behavioral best response can be established following standard assumptions. Next, we define a mean-field equilibrium. 
\begin{definition} [Mean-field equilibrium]
A strategy  profile $a$  generates a mean-field equilibrium if for every player $i,$ the strategy $a_i$ of $i$ is best-response to the others' strategies  $$a_i\in BR_i^{\lambda}(a_{-i}),$$  and it generates a consistent  distribution.
\end{definition}

The existence of mean-field equilibria is not a trivial task. Sufficiency conditions for existence of equilibria can be obtained using fixed-point theory. To do  so, we provide an optimality system for  empathic mean-field-type games.
\subsection{ Dynamic Programming on the space of measures}
Let the expected empathic payoff in terms of the measure $m_t.$  
$$Er_{it}^{\lambda}(s_t, m_t^s, m_t^a, a_t)=\int  r_{it}^{\lambda}(\bar{s}, m_t^s, m_t^a, a_t) m_t^s(d\bar{s})  $$ $$=\hat{r}_{it}^{\lambda}(m_t, a_t)$$ where $\hat{r}^{\lambda}_{it}$ depends only on the measure $m_t$ and the strategy profile $a_t.$ Similarly one can rewrite the expected value of the terminal payoff as
$$Eg_{iT}^{\lambda}(s_T, m_T^s)=\int g_{iT}^{\lambda}(\bar{s}, m_T^s)m_T^s(d\bar{s})=  \hat{g}_{iT}^{\lambda}(m_T^s).$$
\begin{proposition} \label{prov1}
On the space of measures, one has a deterministic dynamic game problem over multiple stages. Therefore a dynamic programming principle (DPP) holds:
$$\left\{ \begin{array}{c}
\hat{v}_{it}^{\lambda}(m_t^s)=\sup_{a'_i}\left\{ \hat{r}_{it}^{\lambda}( m_t,a'_{it},a_{-i,t}) \right.\\  \left. \quad \quad +\hat{v}_{i,t+1}^{\lambda}(m_{t+1}^s) \right\}\\
m_{t+1}^s(ds')= \int_s  q_{t+1}(ds' | \  s, m_t^s, m_t^a, a_t) m_t^s(ds)
\end{array}\right.
$$
\end{proposition}

This optimality system extends the works in \cite{jova82,jova88,b1,marriage,book} to the mean-field-type game case. Note, however that one cannot directly use DPP  
with the state $(s, Em^s, Em^a)$ because of non-Markovian structure. It turns out that one can map the state dynamics to the measure dynamics, and the measure should   be  the state of the DPP. 
\begin{proposition} \label{refpropo}
Suppose a sequence of real-valued function $\hat{v}_{it}^{\lambda},\  t\leq T$ defined on the set of probability measures over $S$ is satisfying the DPP relation above. Then $\hat{v}^{\lambda}_{it}$ is the value function on $P(S)$ starting from $m_t=m.$ Moreover if the supremum is attained  for some $a^*_i(.,m),$ then the best response strategy is in (state-and-mean-field) feedback form. The equilibrium payoff is $$R_i^{\lambda}(a^*)=\hat{v}_{i0}^{\lambda}(m_0).$$
\end{proposition}

Proposition \ref{refpropo} provides a sufficiency  condition for best-response strategies in terms of $(s, m_t^s).$
The proof is immediate and follows from the verification theorem of DPP in deterministic dynamic games.

\subsection{Special cases}

\subsubsection{ Finite state space }
Suppose that the state space $S$ and the action spaces are nonempty and finite. Let the state transition  be $$  P(s_{t+1}=s'\ | \ s_t, m_t^s, m_t^a, a_t)= q_{t+1}(s' | \ s_t, m_t^s, m_t^a, a_t),$$ DPP becomes
$$\left\{ \begin{array}{c}
\hat{v}_{it}^{\lambda}(m_t^s)=\sup_{a'_i}\left\{ \hat{r}_{it}^{\lambda}( m_t^s,  a'_{it},a_{-i,t}) \right.\\  \left. \quad \quad +\hat{v}^{\lambda}_{i,t+1}(m_{t+1}^s) \right\}\\
m_{t+1}^s(s')= \sum_{s\in S} q_{t+1}(s' | \  s, m_t^s, m_t^a, a_t) m_t^s(s)
\end{array}\right.
$$
\begin{proposition} \label{prov2}
A pure mean-field equilibrium may not exist in general. By extending the action space to the set of probability measures on $A$ and the functions $\hat{r}_{it}^{\lambda}, \hat{g}_{iT}^{\lambda}, q_{t+1}$  one gets the existence of mean-field equilibria in behavioral (mixed) strategies.
\end{proposition}

\subsubsection{ Continuous state space }
Consider the state dynamics  
$$s_{t+1}=s_t+ b_{t+1}(s_t, m_t^s, m_t^a, a_t,\eta_{t+1})$$ where $\eta$ is a random process. The transition kernel of $s_{t+1}$ given $s_t, m_t^s, m_t^a, a_t$  is 
$$q_{t+1}(ds' | \  s_t, m_t^s, m_t^a, a_t)=$$ $$ \int_{\eta} P(ds'\ni s_t+ b_{t+1}(s_t, m_t^s, m_t^a, a_t,\eta)) \mathcal{L}_{\eta_{t+1}}(d\eta)$$
where $ \mathcal{L}_{\eta_{t+1}}(d\eta)$ denotes the probability distribution of $ \eta_{t+1}.$

\subsubsection{ Mean-Field Free case}
If $r_i(s,a,m^s,m^a)=r_i(s,a)$ and $g_i(s,m^s)=g_i(s)$ for every player $i$ then $$\hat{r}_i^{\lambda}(m_t,a_t)=\int_s   {r}^{\lambda}_i(s,a_t) m_t(ds).$$ There exists a function ${v}^{\lambda}_i$ such that 
$$\hat{v}^{\lambda}_i(m_t)=\langle  {v}^{\lambda}_i, m_t\rangle= \int_s   {v}^{\lambda}_i(s) m_t(ds),$$   ${v}^{\lambda}_i(s)$ is a mean-field free function. 
In this case, the mean-field-type dynamic programming  reduces to 
$$
{v}_{it}^{\lambda}(s)=\sup_{a'_{it}}H_i^{\lambda}(s, a'_{it},a_{-i,t}),
$$
where the Hamiltonian is
$$ H_i^{\lambda} ={r}_{it}^{\lambda}( s, a'_{it},a_{-i,t}) +\int_{s'} {v}^{\lambda}_{i,t+1}(s')q_{t+1}(ds' |   s_t,  a'_{it},a_{-i,t})$$

We retrieve the classical Bellman operator in the mean-field-free case.
\subsection{ Empathy-Altruism reduces payoff inequality gap}

The following result holds:
\begin{proposition}  \label{prov3}
Empathy-altruism reduces equilibrium payoff inequality gap and improves fairness. As $\lambda$ increases towards $1$ the equilibrium payoff gap between players  $i$ and $j,$ $\hat{v}_{i0}^{\lambda}-\hat{v}_{j0}^{\lambda}$ decreases.
\end{proposition}

\subsection{ Variance Reduction Problem over a Network}\label{sec:example} 
Consider a common state dynamics between  players represented by a stochastic difference equation.
The material cost functional of player $i$  is
	
\begin{equation} \begin{array}{lll} 
L_i(a)= q_{iT}s^2_T+\bar{q}_{iT}(E[s_T])^2\\ \qquad\quad +\sum_{t=0}^{T-1} q_{it} s^2_t+\bar{q}_{it} (E[s_t])^2 +c_{it} a^2_{it}.
\end{array}\end{equation}
which is composed of a terminal material cost  $q_{iT}s^2_T+\bar{q}_{iT}(E[s_T])^2$ and a running material cost of  $q_{it} s^2_t+\bar{q}_{it} (E[s_t])^2 +c_{it} a^2_{it}.$ The coefficients $q_i,\bar{q}_i, c_{it}$ are assumed to be positive real numbers.

The players interact through the common state $s$ which influences the material cost function. This is a dynamic mean-field-type game where the players are not necessarily indistinguishable because the coefficients $q_i,\bar{q}_i, c_i, b_i,$ may be different from one player to another. Moreover, each player $i$ influences  the mean-field term $E[s]$  through its control $a_i.$  In this model, the contribution of a single player (say $i$)  in the mean-field term $E[s]$ cannot be neglected. 
Let $\mathcal{N}_i$ be the set of players that are neighbors of player $i.$ Player $i$ is empathic-altruistic towards her neighbors.
The empathic cost functional of $i$ is $$L_i^{\lambda}(a)=L_i(a)+\sum_{j\in \mathcal{N}_i\backslash \{i\}} \lambda_{ij}L_j(a).$$
We have omitted the term that is not controlled by $i:$  $$\sum_{j\in \mathcal{N}_i\backslash \{i\}}\lambda_{ij} c_{jt} a_{jt}^2$$
  \begin{eqnarray}
\label{LQ00game1type2difference}   
\left\{ 
\begin{array}{lll} \inf_{a_i\in \mathcal{A}_i}  E [L_i^{\lambda}(a_1,\ldots, a_n) ]\
\ 
\displaystyle{\mbox{ subject to }\ }\\
s_{t+1}=\left\{\alpha s_t+\bar{\alpha}Es_t+\sum_{j=1}^n b_j a_{jt}\right\}+\sigma W_{t},\\
s_0\sim \mathcal{L}(S_0),\ \quad \ E[S_0]=m_0  \\ 
 \end{array}
\right.
\end{eqnarray}
given the strategy $(a_j)_{j\neq i}$ of the others' players.

Let $t\in \{0,\ldots, T-1\}$ be the time step,
$ q_{jt}^{\lambda}=\sum_{i\in \mathcal{N}_j}[q_{jt}+\lambda_{ji}q_{it}]\geq 0,\ (q_{jt}^{\lambda}+\bar{q}_{jt}^{\lambda})\geq 0, \ c_{jt}>0,$  and given linear state-and-mean-field feedback of the other players,  the problem (\ref{LQ00game1type2difference}) has  a unique best-response  of player $i$ and it is given by

\begin{equation} \left\{\begin{array}{lll} 
{a}_{it}^{\lambda}=  \eta_{it}(s_t-Es_t)+\bar{\eta}_{it} Es_t,  \\
 \eta_{it}=-\frac{[ \alpha b_i {\beta}_{i,t+1}+b_i{\beta}_{i,t+1}\sum_{j\neq i }b_j\eta_{jt}]  }{{c}_{it}+b_i^2{\beta}_{i,t+1}},\\
\bar{\eta}_{it} =-\frac{b_i{\gamma}_{i,t+1}  (\alpha+\bar{\alpha}+\sum_{j\neq i}b_j\bar{\eta}_{j,t})}{ c_{it}+b_i^2{\gamma}_{i,t+1}},\\
{\beta}_{it}=q_{it}^{\lambda} +{\beta}_{i,t+1}\{ \alpha^2+2\alpha\sum_{j\neq i }b_j\eta_{jt} +[ \sum_{j\neq i }b_j\eta_{jt} ]^2\}\\ -  \frac{[ \alpha b_i {\beta}_{i,t+1}+b_i{\beta}_{i,t+1}\sum_{j\neq i }b_j\eta_{jt}]^2  }{{c}_{it}+b_i^2{\beta}_{i,t+1}}  \\
{\beta}_{iT}= q_{iT}^{\lambda}\geq 0\\
{\gamma}_{it}=({q}_{it}^{\lambda}+\bar{q}_{it}^{\lambda})+{\gamma}_{i,t+1}(\alpha+\bar{\alpha}+\sum_{j\neq i}b_j\bar{\eta}_{jt})^2\\ -\frac{(b_i{\gamma}_{i,t+1}  (\alpha+\bar{\alpha}+\sum_{j\neq i}b_j\bar{\eta}_{jt}))^2}{ c_{it}+b_i^2{\gamma}_{i,t+1}}\\
{\gamma}_{iT}=  q_{iT}^{\lambda}+\bar{q}_{iT}^{\lambda}\geq 0\\
\end{array}\right. \end{equation}
 and the best response cost of player $i$ is
$$
E[ L_i^{\lambda}({a})]= E{\beta}_{i0}(s_0-Es_0)^2+{\gamma}_{i0}(Es_0)^2+ \sum_{t=0}^{T-1}{\beta}_{i,t+1}\sigma^2.$$

We examine the effect of $\lambda$ on the mean state $Es_t.$ Let the real numbers $\alpha,\ \bar{\alpha}, b_i$ and $ (Es_0)$ be nonnegative.
$$Es_{t+1}^{\lambda}= (Es_0)\prod_{k=0}^t [\alpha+\bar{\alpha}+\sum_{i=1}^n b_i \bar\eta_{ik}].$$ It follows that $\gamma^{\lambda}\geq 0$ increases with $\lambda$ and the coefficient $ \bar\eta_{ik} $ decreases with $\lambda.$ We conclude that the  empathy-altruism parameter  helps to  lower the mean state while helping the others in their variance reduction  problem.

\section{Experimental Setup} \label{sec:experiment}
This section presents an experimental evidence of psychological factors in  users' behaviors for data forwarding in Device-to-Device (D2D) communications. 
 \subsection{Reciprocity in Packet Forwarding in D2D Communications}
\begin{figure}[htb]
\includegraphics[scale=0.2]{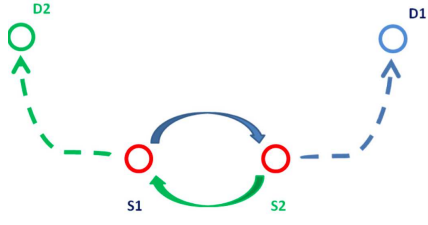}\\ \caption{Forwarding the data of the other nodes. S1 needs S2 to forward its data to D1. S2 needs S1 for forwarding its data to D2. The reciprocity between S1 and S2 helps to improve  connectivity in the  D2D network.}
\label{fig:forward}
\end{figure}

 The explosion of wireless applications creates an ever-increasing demand for more radio spectrum.  The presence of Device-to-Device -enabled mobile users defines an extended network coverage and its co-existence with device-to-infrastructure networks is not without challenges.  In this context, each device can move independently, and will therefore change its links to other devices frequently due to connectivity issues. Relay-enabled wireless device may be requested to forward traffic unrelated to its own use, and therefore be a temporary router or a relay. 
If the receiving device can also play the role of relay then it will forward the data to the next hop after sensing the channel again. For given routing path, the data need to be forwarded at each intermediary hop until the end destination. The intermediary nodes are relays or regular nodes that are willing to forward. 
If many nodes are participating in the forwarding process, every node can benefit from that service, and hence it is a public good, which we refer to as {\it mobile crowdforwarding}. By analogy with crowdfunding, crowdsourcing, crowdsensing, the concept of mobile crowdforwarding consists to call for contributors (mobile devices) who are  willing to forward data in mobile ad hoc networks by means of incentive schemes. However, most of the current smart devices are battery-operated mobile devices that suffer from a limited battery lifetime. Hence, a user who is forwarding a data needs also to balance with the remaining energy by limiting the energy consumptions. When decision-makers are optimizing their payoffs, a dilemma arises because individual and social benefits may not coincide. Since nobody can be excluded from the use of a public good, a user may not have an incentive to forward the data of others. One way of solving the dilemma is to give more incentive to the users. It can be done by slightly changing the game, for example, by adding a second stage in which a reward (fair) can be given to the contributors (non-free-riders).  
 Consider $n$ transmitter-receiver  pairs in a  Device-to-Device  (D2D) communication, with $n\geq 3.$ Player $i$'s action space is $\{nF,F\},$ where $F$ means the player is participating in the forwarding process of the other players' data, and $nF$ refers to not forwarding. There is a need for a critical number $m_*\geq 2$ of participants for the connectivity of  D2D communication networks. In this context it is natural to include a cost sharing within the temporary coalition of nodes. The nodes need the coalition of hops (relays) in order to disseminate information (Fig. \ref{fig:forward}).
 
 The material payoff with sharing cost is 
 
 $$
  r_i=\left\{ \begin{array}{cc}
\prod_{k=1}^{d_i}\ind_{\{SINR_{h_{k-1}h_{k}}\geq \beta_{h_k} \}}   & \mbox{if}\  m> m_* , \ a_i=nF\\
  \prod_{k=1}^{d_i}\ind_{\{SINR_{h_{k-1}h_{k}}\geq \beta_{h_k} \}}  -\frac{m_*}{m}\alpha &  \mbox{if}\  m> m_*,  \ a_i=F\\
   0  & \mbox{if}\  m<m_*,  \ a_i=nF\\
    -\frac{m_*}{m}\gamma   &  \mbox{if}\  m<m_*,  \ a_i=F\\
 \end{array}\right.
 $$
 
 where  $\alpha>0,\ \gamma>0, \ m:= \sum_{j=1}^n \ind_{\{ a_j=1\}}$ and $t_i=h_0,\ h_0h_1,\ldots, h_{l-1}h_{d_i}$ a multihop path from the transmitter of $i$ to the end-to-end destination $d_i.$
 We denote by $p_i:=\prod_{k=1}^{d_i}\ind_{\{SINR_{h_{k-1}h_{k}}\geq \beta_{h_k} \}}.$
 
 \subsection*{Nash equilibria of the game with material payoff}
 If players are only interested in their material payoff, it  leads to the situation where no one would participate in the forwarding of everyone else data.  This is an equilibrium because it is not beneficial to forward the data when no else is forwarding. Moreover the deviant to  pay the cost $-m_*\alpha$ as a single deviator.
 
  \subsection*{Nash equilibria of the game with empathy }
     \subsubsection*{One single deviant does not induce a big degradation}
  If $\lambda_{ij}>0$ for $i$ and $j$  then the  empathetic payoff of $i$ with $a_i=F$ when the number of cooperators exceeds $m_*+1$  is 
  $$
  p_i- \frac{m_*}{m}\alpha +\sum_{j\in \mathcal{N}_{i1}\backslash \{i\} }  \lambda_{ij}(p_j-\frac{m_*}{m}\alpha) +\sum_{j\in \mathcal{N}_{i0} }  \lambda_{ij}p_j
  $$
  and the empathetic payoff for $a_i=nF$ becomes 

    $$
  p_i+\sum_{j\in \mathcal{N}_{i1} }  \lambda_{ij}(p_j-\frac{m_*}{m}\alpha) +\sum_{j\in \mathcal{N}_{i0}\backslash \{i\} }  \lambda_{ij}p_j
  $$
  In this case  a single deviant does not induce big degradation in the payoff.
  
   \subsubsection*{One single deviant limits the performance of the network}
  If $\lambda_{ij}>0$ for $i$ and $j$  then the  empathetic payoff of $i$ with $a_i=F$ when the number of cooperators is $m_*$  is 
  $$
  p_i- \frac{m_*}{m}\alpha +\sum_{j\in \mathcal{N}_{i1}\backslash \{i\} }  \lambda_{ij}(p_j-\frac{m_*}{m}\alpha) +\sum_{j\in \mathcal{N}_{i0} }  \lambda_{ij}p_j
  $$
  and the empathetic payoff for $a_i=nF$ (i.e. $m=m_*-1 < m_*$) becomes 

    $$
  0+\sum_{j\in \mathcal{N}_{i1} }  \lambda_{ij}(p_j-\frac{m_*}{m}\alpha) +\sum_{j\in \mathcal{N}_{i0}\backslash \{i\} }  \lambda_{ij}p_j
  $$
  
  If $p_i- \frac{m_*}{m}\alpha > 0$  then it is better to cooperate because that voice will bring the number of cooperators back to $m_*$ and  the user can take the advantage of the public good.
  
  \subsection*{Nash equilibria of the game with reciprocity}
  If there are enough cooperators, the kindness function yields $ -\frac{m_*}{2m}\alpha $  when $a_i=nF$ and $ +\frac{m_*}{2m}\alpha $ when  $a_i=F.$  Similarly, if the number of cooperators is below $m_*$ then
the kindness function yields $ -\frac{m_*}{2m}\gamma $  when $a_i=nF$ and $ +\frac{m_*}{2m}\gamma$ when  $a_i=F.$ The same methodology determines the sign of the reciprocal perceived kindness. A configuration with $m^*$ cooperators does induce a positive payoff to the users thanks to the kindness and reciprocity of the some of  the others and maintain the public good.
  Thus, altruism and specially reciprocity of the nodes matters in the forwarding process. This is practically observed in the experiment below.

\subsection{Experimental Setup: Empathy and Cooperation}
\begin{figure}[htb]
\includegraphics[scale=0.3]{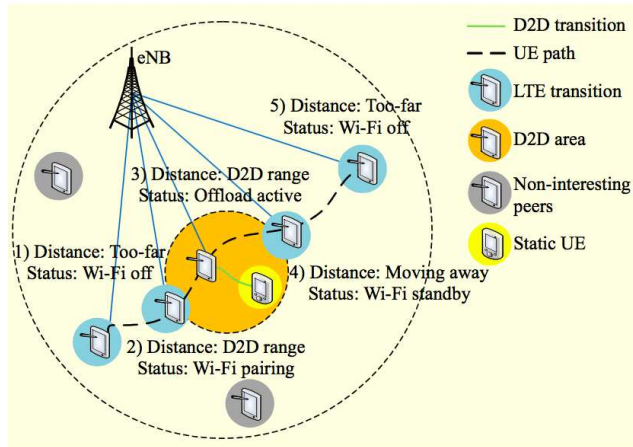}\\ \caption{D2D and WiFi-Direct enabled technology area}
\label{fig:d2dmobility}
\end{figure}

In order to understand the effect of empathy on the behavior of people choice behind the machine, we have conducted an experimental test at NYUAD Learning \& Game Theory Laboratory. We consider a sample population of 47 people carrying wireless devices with  19 men and 28 women, with different cultures, and nationalities and  from 18 to 40 years old. Participants include engineers, psychologists,  students, non-students and professional staff members. In order to quantify the degree of empathy, a multidimensional index measure (Interpersonal Reactivity Index \cite{davis1,davis2}) is used for each member of the population. A participant can move around and may be within a D2D-enabled area if there is a device within a certain range as illustrated in Figure \ref{fig:d2dmobility}. In order to setup a D2D communication network a crucial step is the approval from the users: their decision to cooperate or not in forwarding the data of other devices.

Each person carrying a mobile device was invited to fill a form on its choice in the forwarding dilemma when  facing different configurations. Due to the randomness in wireless channel communications, the forwarding problem  becomes a game under uncertainty.
In the   forwarding game,  the realized payoffs are influenced by the actions of the wireless devices and a random variable representing the channel state. Such games
 are called Random Matrix Forwarding Games. Given random payoff matrices, the question arise as what is meant by playing the random matrix game (RMG, \cite{rmg}) in an optimal way. Because now the actual payoff of the game depends not only on the action profile picked by the wireless devices but also on the sample point of realized state of the channel. Therefore the devices cannot guarantee themselves a  certain payoff level. The wireless devices will have to gamble depending on the channel state. The question of how one gambles in an optimal way needs to be defined. Different approaches have been proposed: 
expectation approach, 
 variance reduction,
mean-variance approach, 
 multi-objective approach.
The signal-to-interference-plus-noise ratio (SINR) for transmission from node S1 to S2 is given by $SINR_{S1S2}= \frac{p |h_{12}|^2}{N_0+I_{S2}},$
where $p>0$ is the transmission power from $S1,$  $ N_0>0$ is the background noise, $h_{S1S2}$ is channel state between S1 and S2, and $I_{S2}\geq  0$ is the interference at the receiver in S2. 
The notation $\ind_{\{ SINR_{S1S2}\geq \beta\}}$ indicates the indicator function on the event $\{ SINR_{S1S2}\geq \beta\},$ i.e., it is equal to $1$ if $SINR_{S1S2}\geq \beta$ and $0$ otherwise.
Let 
$$m_{11}=\ind_{\{ SINR_{S1S2}\geq \beta\}}.\ind_{\{ SINR_{S2D1}\geq \beta\}}, $$ $$n_{11}= \ind_{\{ SINR_{S2S1}\geq \beta\}}.\ind_{\{ SINR_{S1D2}\geq \beta\}},$$
$$n_{12}=\ind_{\{ SINR_{S2S1}\geq \beta\}}.\ind_{\{ SINR_{S1D2}\geq \beta\}},$$
$$m_{21}=\ind_{\{ SINR_{S1S2}\geq \beta\}}.\ind_{\{ SINR_{S2D1}\geq \beta\}}.$$
Since $h=(h_{S1S2},h_{S2S1},h_{S1D2},h_{S2D1})$ is a random vector, the coefficients $m_{11},n_{11},n_{12},m_{21}$ are random. This leads to a  random matrix forwarding game between  wireless devices S1 and S2 as described in Table \ref{table1v2}.

\begin{table}[htb]
  \centering
\begin{tabular}{|c|c|c|}
  \hline
   S1$\backslash$ S2 & $F$ & $nF$ \\ \hline
  $F$ & $(m_{11}-c_1, n_{11} -c_2)$&  $(-c_1, n_{12})$ \\ \hline
  ${nF}$ & $(m_{21}, -c_2)$ & $(0,0)$ \\  \hline
\end{tabular} \caption{Random matrix forwarding game.}\label{table1v2}
\end{table}
 We describe below the expectation approach. It consists to replace the coefficients of payoff matrix by the corresponding mathematical expectation where the expectation is taken with the respect to $h.$ We denote by $a_{ij}:=\mathbb{E}\left[ m_{ij}\right]$ and $b_{ij}:=\mathbb{E}\left[ n_{ij}\right].$ Then expected payoff matrix is given by Table \ref{table1v3}.
 
\begin{table}[htb]
  \centering
\begin{tabular}{|c|c|c|}
  \hline
  S1$\backslash$ S2 & F & nF \\ \hline
  F & $(a_{11}-c_1, b_{11} -c_2)$&  $(-c_1,b_{12})$ \\ \hline
  nF & $(a_{21}, -c_2)$ & $(0,0)$ \\
  \hline
\end{tabular} \caption{Expected matrix forwarding game.}\label{table1v3}
\end{table}

\subsubsection*{No empathy implies no network in most interesting cases}
We analyze the normal form game of Table \ref{table1v3}. If  $a_{11}-c_1< a_{21}$ then the row player will not forward, and hence the column player as well. This leads to  Nash equilibrium strategy $(nF,nF).$ If  $a_{11}-c_1> a_{21}$ then the row player will  forward, and the column player will forward if   $b_{11} -c_2>b_{12}$ leading to $(nF,nF)$,  else if $b_{11} -c_2<b_{12}$ then the equilibrium is $(F,nF).$ Similar reasoning can be conducted by inverting the roles. Thus, when taking into consideration the power-limited of the mobile devices, the classical material payoff analysis leads to  the outcome $(nF,nF)$ i.e., non-cooperation between the mobile users, and no forwarding implies no network. 

\subsubsection*{Effect of empathy on  the forwarding decision}
Now we involve possible empathetic situations. Two  contexts were available in the game situation. The first context is a situation where the two persons involved in the game are friends. The second context is when they do not know each other (they meet for example during in their way in the public transportation, and do not have an a priori relationship. The empathy measure used for the experiment is the so-called 
the  Interpersonal Reactivity Index (IRI) which  comprised of four scales:  empathic concern (EC), perspective taking (PT), fantasy scale (FS) and personal distress (PD).
\begin{itemize}\item The EC scale aims to assess the affective outcomes, the tendency to experience other-oriented feelings and the response to distress in others with the reactive response of sympathy and compassion. \item   The PT scale aims to measure the process of role taking, the tendency to adopt the psychological points of view of others. \item  The PD scale  demonstrates an affective outcome, and is designed to tap ones' own feelings of personal unease and discomfort in reaction to the emotions of others. 
 \item the FS aims to measure the tendency to transpose oneself into feelings and actions of fictitious characters.
\end{itemize}

\subsubsection{Procedure of the experiment}

Participants were run individually, although they were led to believe that another person  was also taking part.  The experimenters explained to each participant that the study involved two participants, and that they were being placed temporarily at different places. The experimenters then escorted the participant to the NYUAD Learning \& Game Theory Laboratory left her alone to read a 
written instruction that allows us the measure its empathy subscales, followed by another instruction on the packet forwarding and participation into D2D technology.   The test also distinguishes the 
gender of the participant, in order to make a refined study with several types and subpopulations. After participants read the questionnaire (see Table \ref{tab:table1}), the experimenter answered any questions, and informed them that they and the other participant in the session had been randomly assigned to and  the experimenter returned.  If carefully filled, the instructions reveal a significant empathy scale, the latency per question and the decision of the participant in the forwarding game in two different situations: close relationship with other participant that was fictitious in the test or no prior relationship with the participant. All participants have wireless devices that have the capability in enabling WiFi direct and D2D technology when the users decide to do so. They have the possibility in accepting or rejecting  (to enable or to disable)  to forwarding the data of the others.

\subsubsection{Participation to the experiment}
 In the men population  only  two questions have  been left in IRI, with a 99.63\% of responsiveness to the four different scales. In women population we had three questions that have been left in IRI, with a 99.62\% of responsiveness to the same four different scales.

\subsubsection{Analysis of the experimental data}
\pgfplotsset{
  compat=newest,
  xlabel near ticks,
  ylabel near ticks
}
\begin{tikzpicture}[font=\small]
    \begin{axis}[
      ybar interval=0.3,
      bar width=2pt,
      grid=major,
      xlabel={Empathy Scale Quality: Perspective Taking (PT)},
      ylabel={Total Score of Answers},
      ymin=0,
      ytick=data,
      xtick=data,
      axis x line=bottom,
      axis y line=left,
      enlarge x limits=0.1,
      symbolic x coords={awful,bad,average,good,excellent,ideal},
      xticklabel style={anchor=base,yshift=-0.5\baselineskip},
    ]
           \addplot[fill=yellow] coordinates {
               (awful,4) (bad,48)  (average,50)   (good,19)  (excellent,4) (ideal, 30)
  }; \addplot[fill=white] coordinates {
      (awful,4)
       (bad,16)
        (average,24)
         (good,21)
        (excellent,4) (ideal, 30)      
      };

\legend{ Women, Men}
    \end{axis}
  \end{tikzpicture}

\begin{tikzpicture}[font=\small]
    \begin{axis}[
      ybar interval=0.3,
      bar width=2pt,
      grid=major,
      xlabel={Outcomes of the Data Forwarding Game },
      ylabel={Number of "yes"},
      ymin=0,
      ytick=data,
      xtick=data,
      axis x line=bottom,
      axis y line=left,
      enlarge x limits=0.1,
      symbolic x coords={FF,FnF,nFF,nFnF,other},
      xticklabel style={anchor=base,yshift=-0.5\baselineskip},
    ]
      \addplot[fill=white] coordinates {
        (FF,19)
        (FnF,16)
        (nFF,4)
        (nFnF,16) (other, 10)
      };
      \addplot[fill=yellow] coordinates {
        (FF,10)
        (FnF,1)
        (nFF,5)
        (nFnF,13) (other, 8)
      };
\legend{Women, Men}
    \end{axis}
  \end{tikzpicture}

\begin{itemize}
\item Women population:
\begin{tabular}{cc|c|c|c|c|l}
\cline{3-4}
& & \multicolumn{2}{ c| }{Player I} \\ \cline{3-4}
& & F& nF  \\ \cline{1-4}
\multicolumn{1}{ |c  }{\multirow{2}{*}{Player II} } &
\multicolumn{1}{ |c| }{F} & 19 & $ 16$      \\ \cline{2-4}
\multicolumn{1}{ |c  }{}                        &
\multicolumn{1}{ |c| }{nF} & $4$ & 16    \\
 \cline{1-4}
\end{tabular}

A more refined version of the cooperators among the women population with FnF/nFF outcomes (15/28) is obtained:

\begin{tabular}{cc|c|c|c|c|l}
\cline{3-4}
& & \multicolumn{2}{ c| }{Player I} \\ \cline{3-4}
& & F& nF  \\ \cline{1-4}
\multicolumn{1}{ |c  }{\multirow{2}{*}{Player II} } &
\multicolumn{1}{ |c| }{F} & 10 & $ 1$      \\ \cline{2-4}
\multicolumn{1}{ |c  }{}                        &
\multicolumn{1}{ |c| }{nF} & $5$ & 13    \\
 \cline{1-4}
\end{tabular}

\item Men population:

\begin{tabular}{cc|c|c|c|c|l}
\cline{3-4}
& & \multicolumn{2}{ c| }{Player I} \\ \cline{3-4}
& & F& nF  \\ \cline{1-4}
\multicolumn{1}{ |c  }{\multirow{2}{*}{Player II} } &
\multicolumn{1}{ |c| }{F} &  11& $ 8$      \\ \cline{2-4}
\multicolumn{1}{ |c  }{}                        &
\multicolumn{1}{ |c| }{nF} & $3$ & 13    \\
 \cline{1-4}
\end{tabular}
\end{itemize}
A more refined version of the cooperators among men population with  FnF/nFF outcomes  shows (6/19)  proportion of cooperators.

 \subsection{Observations from the experiment}
 
 \begin{itemize}
 \item 
  Although they read identical notes, we expected that participants who had close relationship  would experience more empathy for the other participant than would participants who do not each other and never met before (either virtually or physically). We checked this expectation with the self-reports of IRI response that participants made after reading an alternative question on what would be their decision if they do not know the other participant. It turns out  that only 1 person (out of 47 people) will change their opinion if the other user is unknown to them.  Thus, both empathy and closiness affect the decision-making of the users.
\item Our second observation is that the experiment exhibits  a strong correlation between the scale of IRI and the choice of  people.  Figure \ref{fig:cooperation} illustrates a relationship with PT scale and percentage of cooperators.

\begin{figure}[htb]
\includegraphics[scale=0.2]{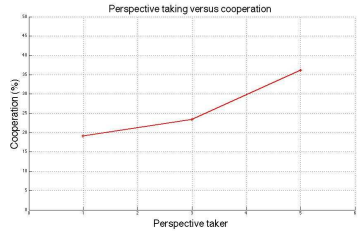}\\ \caption{Impact of positive empathy (PT) in the decision-making of the people}
\label{fig:cooperation}
\end{figure}

Using total probability theorem we obtain 
 $$P(F)=P(F|1)P(1)+P(F|0)P(0), $$
where  $P(F| i)$ is the conditional probability of forwarding the data of the others (cooperation) assuming $i.$ We use the sample statistics to compute the probability to cooperate through the number of occurrences of $F.$ 
\item Deviation to the material payoff outcomes:  What if participants in a one-shot prisoner's dilemma game know before making their decision that the other person has already decided not to forward (defected)?  From the perspective of classic game theory with material payoff, a dilemma no longer exists because of dominating strategy. It is clearly in their best interest to defect too. The empathy-based test predicts, however, that if some of them feel 
empathy for the other, then a forwarding dilemma remains: self-interest counsels  {\it not to forward} (defection); empathy-induced behavior may counsel not.  
Based on the experiment we have look at the outcomes $(F, nF)$ and $(nF,nF)$ from the choices of 47 participants.
Among those not induced to feel empathy, very few (3/47) did not defect in return. Among those induced to feel empathy for the other,  (26/47) did not defect. These experimental results highlight the power of empathy-induced behavior to affect decisions in one-shot forwarding dilemma game.

  \item This experimental test reveals that empathy seems far more effective than most other techniques that have been proposed to increase cooperation in one-shot games.

\end{itemize}

Based on these experimental results, we believe that the idea of 
using (positive) empathy to increase cooperation in a one-shot forwarding dilemma and more generally  in a public good games  should be  explored in more details.

Can we use psychological payoff functions  to explain the behaviors observed in the experiment?

To answer this question we introduce a psychological payoff that is not only self-interested but also other-regarding through the two random variables $\lambda_{1}$ and $\lambda_{2}.$

\begin{table}[htb]
  \centering
\begin{tabular}{|c|c|c|}
  \hline
   S1$\backslash$ S2 & $F$ & $nF$ \\ \hline
  $F$ & $(m_{11}^{\lambda}, n_{11}^{\lambda})$&  $(m_{12}^{\lambda},n_{12}^{\lambda})$ \\ \hline
  ${nF}$ & $(m_{21}^{\lambda},n_{21}^{\lambda})$ & $(0,0)$ \\  \hline
\end{tabular} \caption{Random matrix forwarding game}\label{table1v22}
\end{table}

$$m^{\lambda}_{11}=m_{11}-c_1 +\lambda_1( n_{11} -c_2),\   n^{\lambda}_{11}=\lambda_2(m_{11}-c_1) +n_{11} -c_2$$
$$m^{\lambda}_{12}=-c_1 +\lambda_1 n_{12},\  n^{\lambda}_{12}=-\lambda_2c_1 + n_{12}$$

$$m^{\lambda}_{21}= m_{21}-\lambda_1 c_2,\  \ n^{\lambda}_{21}= \lambda_2 m_{21}- c_2 $$

\begin{itemize}
\item Case 0: In absence of empathy: $\lambda_1=0, \lambda_2=0$  corresponds to the self-regarding payoffs case. The game leads to the outcome (nF,nF) when $m_{11}-c_1 <   m_{21}$ and $n_{11}-c_2 <   n_{12}.$
\item Case 1:  $\lambda_1>0,\ \lambda_2>0 $ 

\begin{itemize}\item FF:
If $m^{\lambda}_{11}\geq m^{\lambda}_{21}$  or  $n^{\lambda}_{11}\geq n^{\lambda}_{12}$ then  the strategy nF is not dominating anymore  and in this case, forwarding  is a good candidate for Nash equilibrium of the psychological  one-shot forwarding  game. In addition, if $m_{ii}\geq 0$ and $\lambda_1\geq  \frac{c_1+m_{21}- m_{11}}{n_{11} }$ and $ \lambda_2  >   \frac{c_2 + n_{12}-n_{11}}{m_{11}} $ then 
full cooperation $(F,F)$ becomes a Nash equilibrium.  
If  $\frac{c_1+m_{21}- m_{11}}{n_{11} }$ and  $ \frac{c_2 + n_{12}-n_{11}}{m_{11}} $ belongs to $(0,1)$ and  then a mixed strategy equilibrium emerges in addition to the pure ones, which explains  the observed variation of  percentages of cooperators depending on the empathy index measured from the experiment. 

\item FnF is an equilibrium if $ m_{12}^{\lambda}\geq 0$ and $n_{12}^{\lambda}\geq n_{11}^{\lambda}.$ This means that  $\lambda_1 n_{12}\geq c_1,\  \lambda_2 m_{11}+ n_{11}-n_{12}-c_2\leq 0,$ i.e., $\lambda_1$ positively high enough and $\lambda_2$ is low.
\item Similarly when $\lambda_1$ is low and  $\lambda_2$ positively high enough then $nFF$ becomes an equilibrium.
\item nFnF is an equilibrium when $ \{m_{12}^{\lambda}\leq 0, n_{21}^{\lambda}\leq 0\}$ which means $\lambda_1 n_{12}\leq c_1,   \lambda_2 m_{21}\leq  c_2.$
\item  If $ \lambda_1\in ( \frac{c_1+m_{21}- m_{11}}{n_{11} }, \frac{c_1}{n_{12} }) $  and  $\lambda_2\in (\frac{c_2 + n_{12}-n_{11}}{m_{11}} ,\frac{c_2}{m_{21}})$ then there are three equilibria: FF, nFnF and a mixed equilibrium.
\item If $ \lambda_1>\frac{c_1}{n_{12} }$ and $\lambda_2 >\frac{c_2}{m_{21}}$ then FF is the unique equilibrium because $F$ is a dominating strategy for both users.
\item If   both users have low empathy $\lambda_1<\frac{c_1+m_{21}- m_{11}}{n_{11} }$ and   $\lambda_2<\frac{c_2 + n_{12}-n_{11}}{m_{11}}$ then $nF$ is a dominating strategy for both users, hence $nFnF$ is an equilibrium.
\end{itemize}
\item Case 2:  $\lambda_1<0,\ \lambda_2<0 :$  Both users have a dominating strategy which is nF.  Then, nFnF is the outcome. 
\item Case 3:  $\lambda_1>0,\ \lambda_2<0: $   Player 2 has a dominating  strategy which is nF. Thus, FnF is the outcome if the empathy of player  $1$ is high enough and $nFnF$ otherwise. At the threshold value  of $\lambda$ such that $m_{12}^{\lambda}=0,$ every partially  mixed strategy profile $ (y \delta_{F}+(1-y)\delta_{nF}, nF)$ with $y\in [0,1]$ is an equilibrium.
\item Case 4:  $\lambda_1<0,\ \lambda_2>0: $  nFF is the outcome if the empathy of $2$ is high enough and $nFnF$ otherwise.
\end{itemize}
\begin{figure*}[htb]
\includegraphics[scale=0.3]{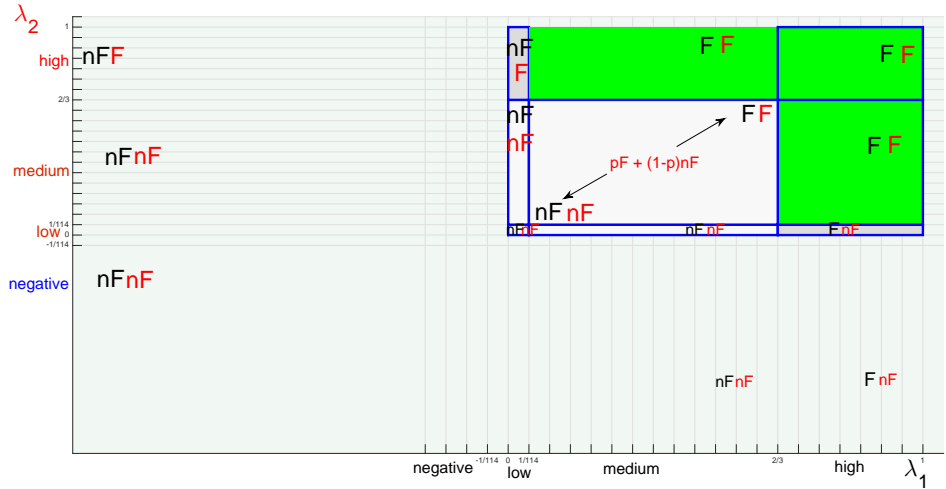}\\ \caption{Outcome based on empathy distribution}
\label{fig:empathyte}
\end{figure*}

When the parameters lead to an equality of payoff, there may be infinite number of (mixed) equilibria. We have omitted these degenerate cases since $\lambda$ will be a continuous  random variable (Figure \ref{fig:empathyte} and table \ref{tab:sum}).
Table \ref{tab:sum} summarizes the outcomes when the entries $m,n$ are non-zero depending on the affective empathy level of the users: negative (spiteful), low (positive), medium, and high when $\frac{c_1+m_{21}- m_{11}}{n_{11} }<  \frac{c_1}{n_{12} }$ and 
$\frac{c_2+n_{21}- n_{11}}{m_{11} }< \frac{c_2}{m_{21} }.$

\begin{table}
\begin{tabular}{|c|c|c|c|c|}
  \hline
 User 1 $\backslash$ User 2&  $\lambda_{2}$  Negative & Low & Medium &  High \\ \hline
 $\lambda_{1}$  {\color{green} High} & { \color{green} F}nF &  {\color{green} F}nF & {\color{green} F F}&   {\color{green} FF}\\
 $\lambda_{1}$  Medium & nFnF & nFnF& {\color{green} FF},nFnF, p{ \color{green}F}+(1-p)nF & {\color{green} FF} \\
  $\lambda_{1}$  Low & nFnF & nFnF& nFnF&  nF{\color{green} F}\\ \hline
   $\lambda_{1}$ Negative & nFnF&  nFnF &  nFnF&nF{\color{green} F}\\   \hline
\end{tabular}

 \caption{Summary of the outcomes. For user 1,   Low empathy means  $\lambda_1\in (0, \frac{c_1+m_{21}- m_{11}}{n_{11} }),$ Medium empathy means $\lambda_1\in (\frac{c_1+m_{21}- m_{11}}{n_{11} },  \frac{c_1}{n_{12} })$ and High empathy means $\lambda_{1}> \frac{c_1}{n_{12} }.$  For user 2, low empathy means  $\lambda_1\in (0, \frac{c_2+n_{21}- n_{11}}{m_{11}}).$ 
 Medium empathy means $\lambda_2\in (\frac{c_2+n_{21}- n_{11}}{m_{11} },  \frac{c_2}{m_{21} })$ and High empathy means means $\lambda_{2}> \frac{c_2}{m_{21} }.$      }\label{tab:sum} 
\end{table}

This experimental test reveals a distribution of empathy across the population  of men and women (see Figure \ref{fig:repartition}). Thus, a natural question is the probability to endup with FF as an outcome when people are drawn from the empathy  population sampling distribution over $[-1,1]^{2}.$

Below we examine the extreme cases with two types: High PT and Se distributed according  to  $(1-\mu, \mu)$ for some $\mu\in (0,1).$
The resulting interaction depends on  the type of the users carrying the wireless nodes. We denote by $PT$  a high level of positive empathy and by $Se$ a user with a very low level of empathy.
\begin{table}[htb]
  \centering
\begin{tabular}{|c|c|}
  \hline
  PT$\backslash$ PT & $F$  \\ \hline
  $F$ & $(m_{11}^1-c_1, m_{11}^2-c_2)^*$ \\
  \hline
\end{tabular}   \ \  \\  \begin{tabular}{|c|c|c|}
  \hline
  PT$\backslash$ Se & $F$  & ${nF}$ \\ \hline
  $F$ & $(m_{11}^1-c_1,m_{11}^2-c_2)$  & $(-c_1,m_{12}^2)$\\
  \hline
\end{tabular}\\ \vspace{1cm}
\begin{tabular}{|c|c|}
  \hline
  Se$\backslash$ PT & $F$  \\ \hline
  $F$ & $(m_{11}^1-c_1,m_{11}^2-c_2)$ \\ \hline
   $\bar{F}$ & $(m_{21}^1,-c_2)$ \\
  \hline
\end{tabular} \ \  \\
\begin{tabular}{|c|c|c|}
  \hline
  Se$\backslash$ Se & $F$ & ${nF}$ \\ \hline
  $F$ & $(m_{11}^1-c_1,m_{11}^2-c_2)$&  $(-c_1,m_{12}^2)$ \\ \hline
  ${nF}$ & $(m_{21}^1, -c_2)$ & $(0,0)$ \\
  \hline
\end{tabular}
 \caption{Random matrix game in strategic form with incomplete information.}\label{table2rmg} 
\end{table}

\begin{itemize}
\item In a PT-PT interaction, the equilibrium structure is to forward whenever the channel is good enough.
\item In a Se-PT or  PT-Se interaction, the selfish wireless node has to compare.   In a Se-PT interaction, if  $m_{11}^1-c_1>m_{21}^1$ then the selfish node  1 will choose $F$ otherwise will not forward (${nF}$).  The equilibrium structure of Se-PT interaction is  \begin{itemize}\item $(F,F)$  if  $m_{11}^1-c_1>m_{21}^1,$ \item $({nF},F)$ if  $m_{11}^1-c_1<m_{21}^1$, \item $(\mbox{any mixed strategy}, F)$  if $m_{11}^1-c_1=m_{21}^1$
\end{itemize}
Similarly, the equilibrium structure of PT-Se interaction is  \begin{itemize}\item $(F,F)$  if  $m_{11}^2-c_2>m_{12}^2,$ \item $(F,{nF})$ if  $m_{11}^2-c_2<m_{12}^2,$ \item $(F,\mbox{any mixed strategy})$  if $m_{11}^2-c_2=m_{12}^2,$ \end{itemize}
\item In a Se-Se interaction, the equilibrium structure is described in a similarly way as in S1-S2 in Table \ref{table1v3}.
\end{itemize}

Note that when we put the pure strategies together in a population context where $(1-\mu)$ fraction of the people are empathic PT and $\mu \in (0,1)$ fraction are selfish nodes, the resulting outcome is well-mixed of 
$(F,F), (F,nF), (nF, F) $ and $(nF, nF),$ which  strengthen the observations of the experiment. It also provides the possibility to observe the Bayesian Hannan set \cite{hannan} or Bayesian coarse correlated equilibria in experimental one-shot games.  

\subsection{Other  empathy subscales}
\begin{figure}[htb]
\centering
\includegraphics[scale=0.5]{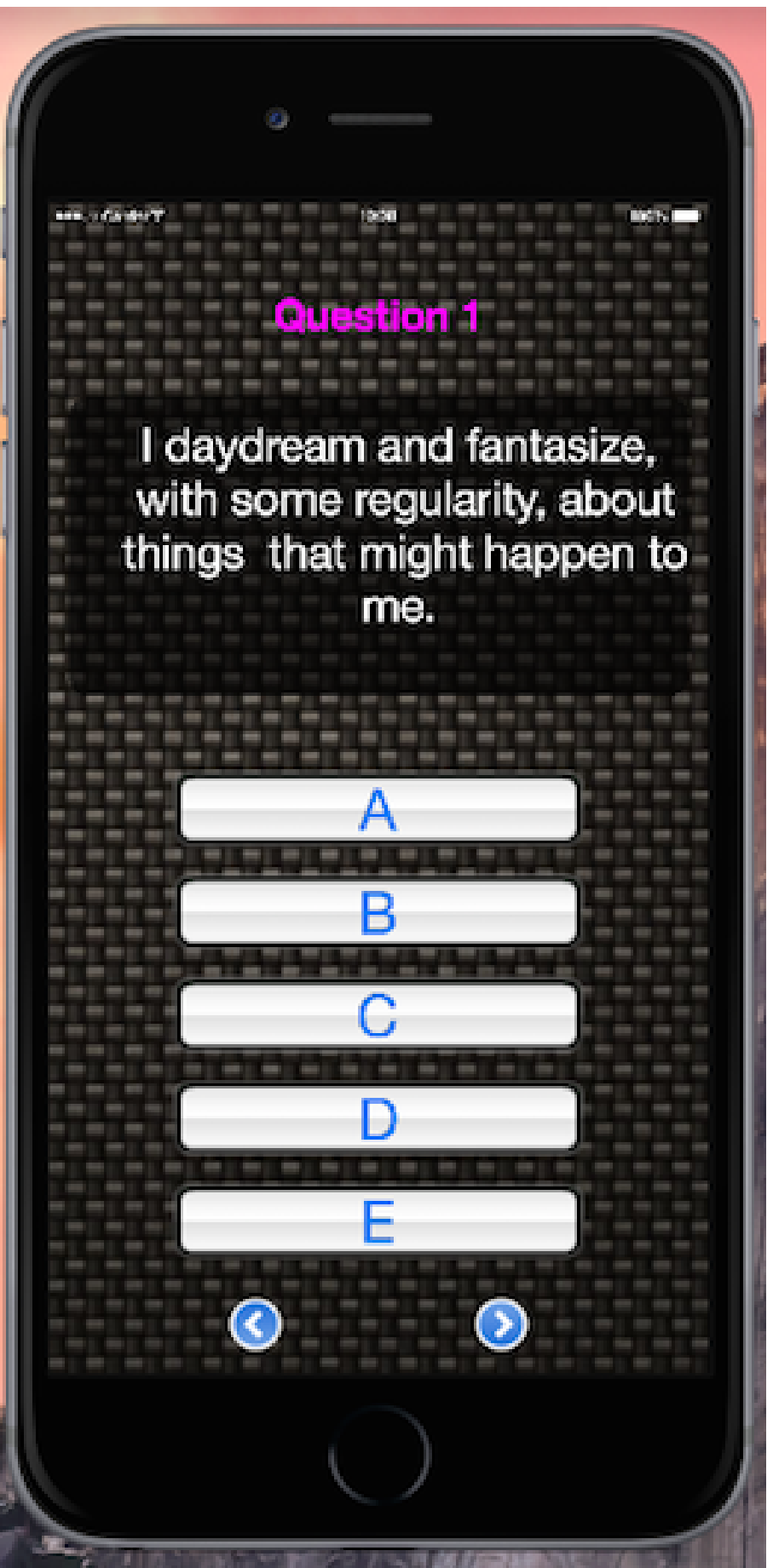}\  \includegraphics[scale=0.5]{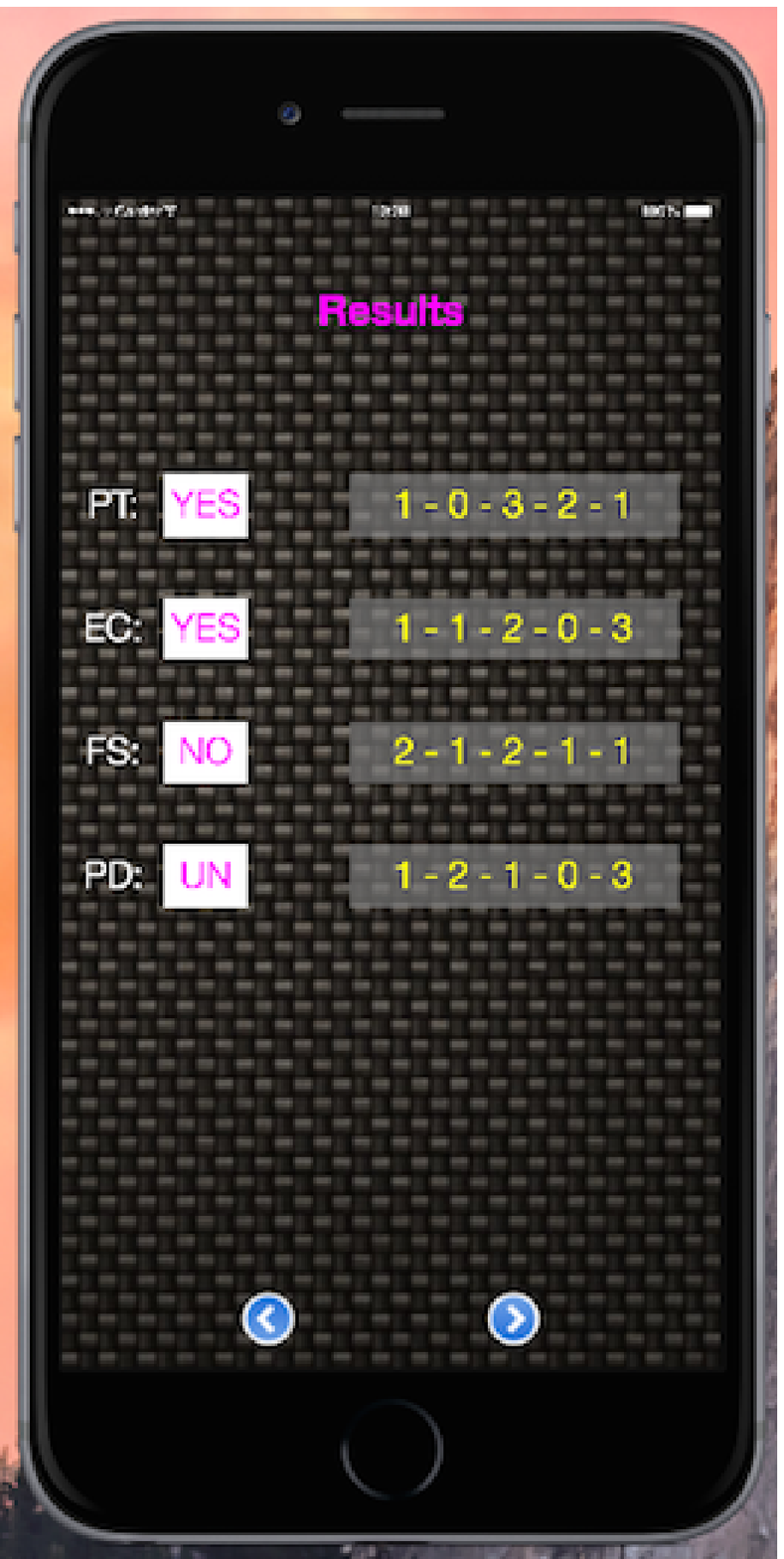}\caption{Empathizer  measures the multidimensional empathy of each participant.}
\label{fig:empathizer2}
\end{figure}

\begin{figure}[htb]
\centering
\includegraphics[scale=0.6]{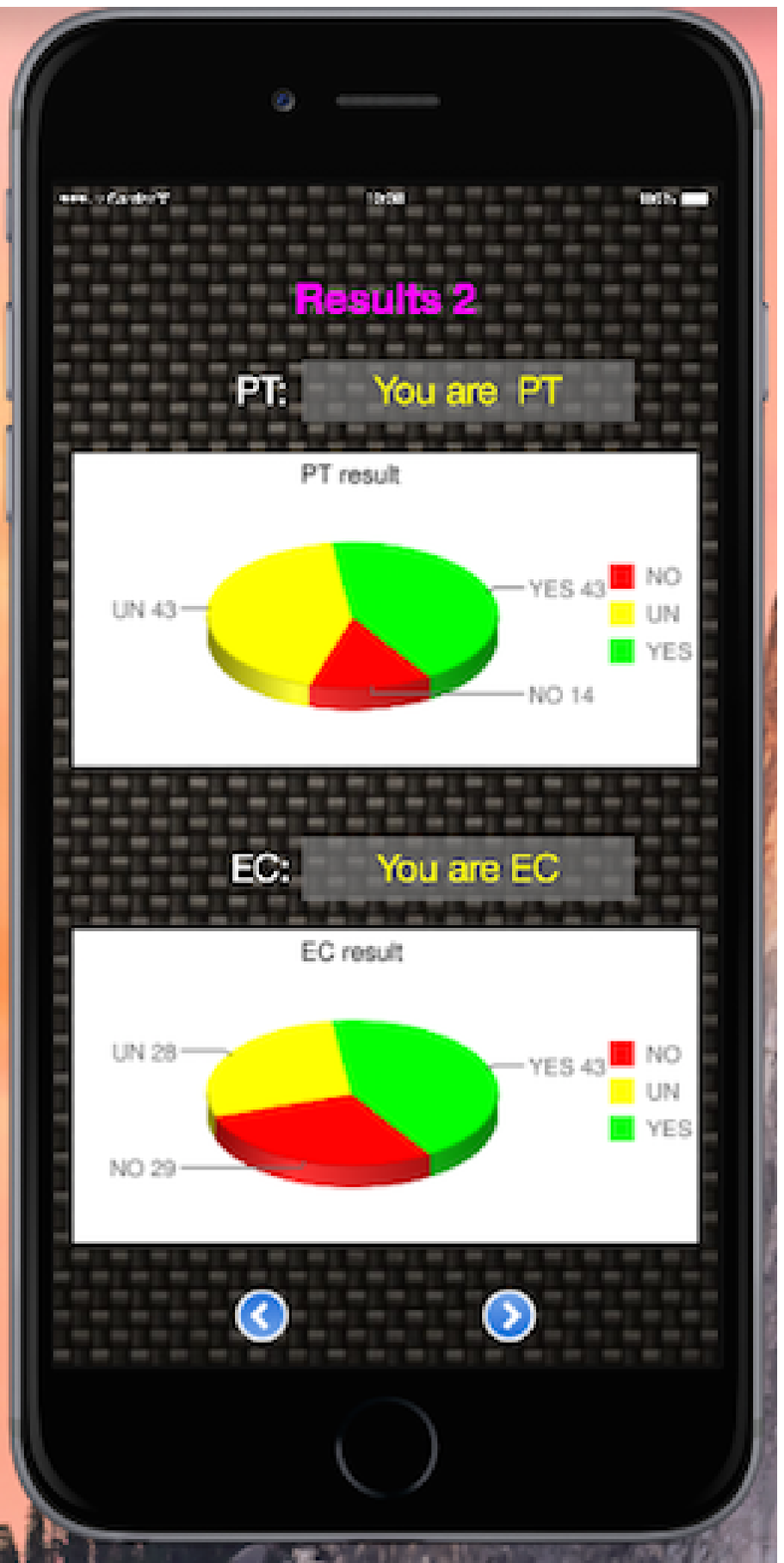}\  \includegraphics[scale=0.6]{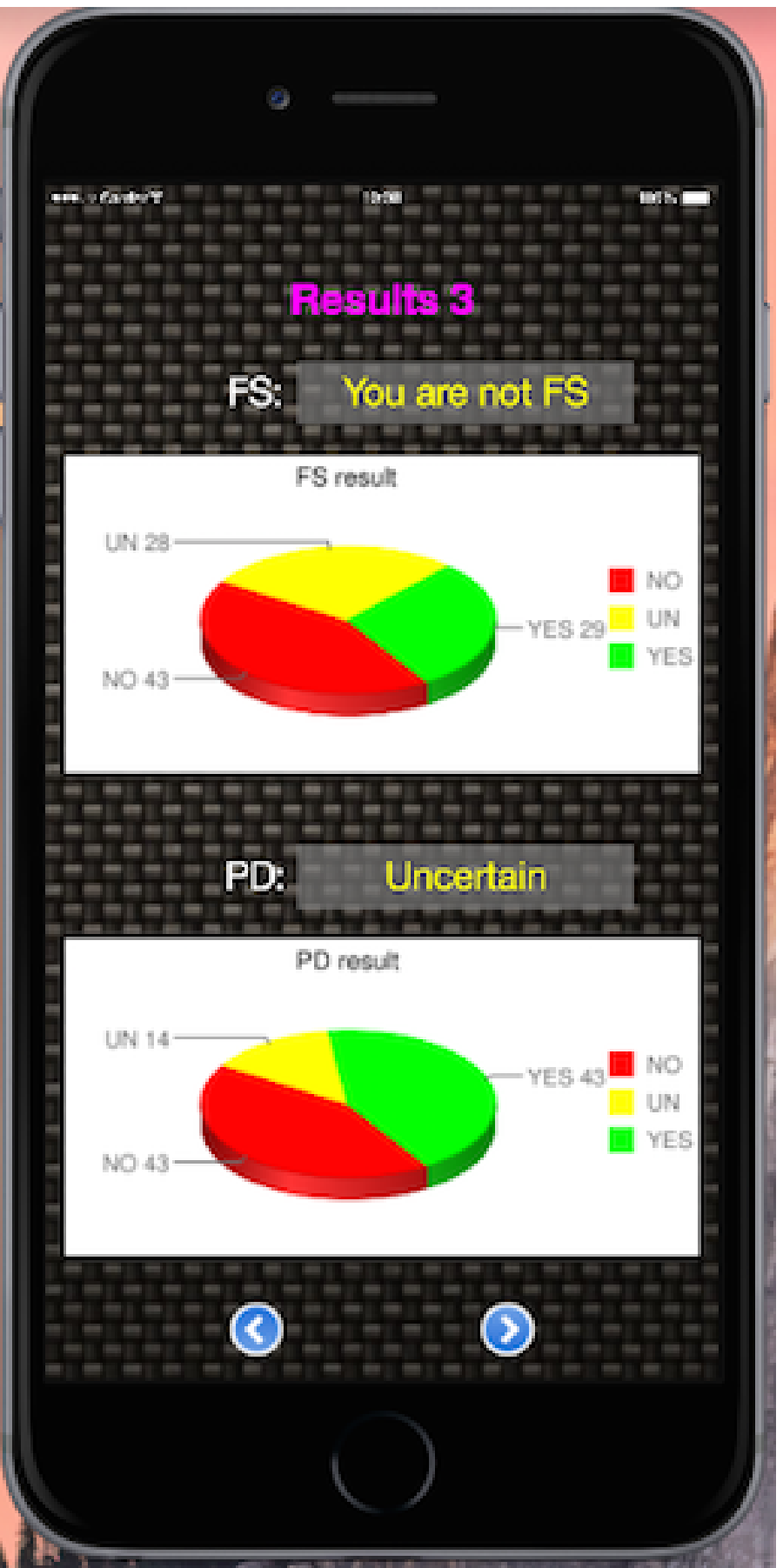}\caption{Empathizer  sample result.}
\label{fig:empathizer3}
\end{figure}

Screen shots of  the Empatizer app are given in Figures \ref{fig:empathizer2} and \ref{fig:empathizer3}.
As expected, the classification is incomplete and the subscales are not statistically independent. They are correlated and possibly overlapping. 
This is represented in Figure \ref{fig:repartition}, Table \ref{table:IRIdistr1}. 

\begin{table}[h]
\centering
\begin{tabular}{|llll|}
\hline 
{\bf Scale Type} & {\bf Women} & {\bf Men}  & {\bf  Total } \\ \hline
PT  & 5 & 3 &  8 \\ \hline
EC     & 2    &  -  &  2  \\ \hline
FS     & 3    &   3 &  6  \\ \hline
PD     & 2    &   4 &  6  \\ \hline

PT - PD    & 1    &   1 &  2  \\ \hline
FS - PD    & 2   &   - &  2  \\ \hline
EC - PD    & -    &   3 &  3  \\ \hline
EC - FS    & -    &   1 &  1  \\ \hline

PT - EC - PD    & -    &   1 &  1  \\ \hline
EC - FS - PD    & 1    &  -  &  1  \\ \hline

PT - EC - FS - PD    & -   &   1 &  1  \\ \hline

Other scale  & 12    &   2 &  14  \\ \hline
  &  1  &   1&  2\\ \hline
Participants  &    28&   19 &  47  \\ \hline

\end{tabular}
\caption[]{ IRI scale distribution across the population}
\label{table:IRIdistr1}
\end{table}

 \def\angle{0}
\def\radius{3}

\def\cyclelist{{"green", "blue","orange","red", "yellow"}}
\newcount\cyclecount \cyclecount=-1
\newcount\ind \ind=-1

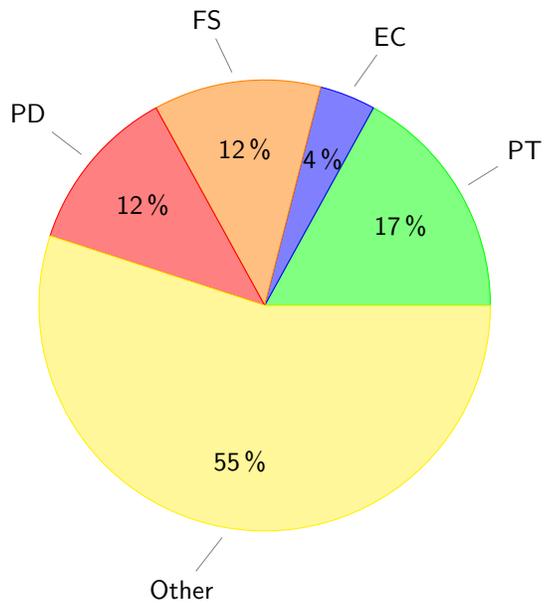
\begin{figure}
\begin{tikzpicture}[nodes = {font=\sffamily}]
  \foreach \percent/\name in {
      17/PT,
      4/EC,
      12/FS,
      12/PD,
      55/Other
    } {
      \ifx\percent\empty\else               
        \global\advance\cyclecount by 1     
        \global\advance\ind by 1            
        \ifnum4<\cyclecount                 
          \global\cyclecount=0              
          \global\ind=0                     
        \fi
        \pgfmathparse{\cyclelist[\the\ind]} 
        \edef\color{\pgfmathresult}         
        \draw[fill={\color!50},draw={\color}] (0,0) -- (\angle:\radius)
          arc (\angle:\angle+\percent*3.6:\radius) -- cycle;
        \node at (\angle+0.5*\percent*3.6:0.7*\radius) {\percent\,\%};
        \node[pin=\angle+0.5*\percent*3.6:\name]
          at (\angle+0.5*\percent*3.6:\radius) {};
        \pgfmathparse{\angle+\percent*3.6}  
        \xdef\angle{\pgfmathresult}         
      \fi
    };
\end{tikzpicture}
\caption{Empathy scale distribution across the population of participants}
\label{fig:repartition}
\end{figure}

In  Table \ref{tab:table1} we have completed the subscales PD, FS, EC.  We observe that both positive and negative correlation between cooperation and empathy subscale can be obtained from the experiment, in particular for PD. This also reveals a kind of spitefulness behavior. However, it is a mixture of several things. for example, if player $i$ is spiteful towards $j$ and $j$ is PT, the resulting outcome is unclear.

\begin{tikzpicture}[font=\small]
    \begin{axis}[
      ybar interval=0.3,
      bar width=2pt,
      grid=major,
      xlabel={Empathy Scale Quality: Empathy Concern (EC)},
      ylabel={Total Score of Answers},
      ymin=0,
      ytick=data,
      xtick=data,
      axis x line=bottom,
      axis y line=left,
      enlarge x limits=0.1,
      symbolic x coords={awful,bad,average,good,excellent,ideal},
      xticklabel style={anchor=base,yshift=-0.5\baselineskip},
    ]
           \addplot[fill=yellow] coordinates {
               (awful,69) (bad,49)  (average,22)   (good,48)  (excellent,18) (ideal, 30)
  };
   \addplot[fill=white] coordinates {
      (awful,36)
       (bad,23)
        (average,27)
         (good,29)
        (excellent,9) (ideal, 30)      
      };

\legend{ Women, Men}
    \end{axis}
  \end{tikzpicture}


\begin{tikzpicture}[font=\small]
    \begin{axis}[
      ybar interval=0.3,
      bar width=2pt,
      grid=major,
      xlabel={Empathy Scale Quality: Personal Distress (PD)},
      ylabel={Total Score of Answers},
      ymin=0,
      ytick=data,
      xtick=data,
      axis x line=bottom,
      axis y line=left,
      enlarge x limits=0.1,
      symbolic x coords={awful,bad,average,good,excellent,ideal},
      xticklabel style={anchor=base,yshift=-0.5\baselineskip},
    ]
           \addplot[fill=yellow] coordinates {
               (awful,16) (bad,32)  (average,26)   (good,29)  (excellent,4) (ideal, 30)
  };
   \addplot[fill=white] coordinates {
      (awful,4)
       (bad,20)
        (average,24)
         (good,17)
        (excellent,26) (ideal, 30)      
      };

\legend{ Women, Men}
    \end{axis}
  \end{tikzpicture}


\begin{tikzpicture}[font=\small]
    \begin{axis}[
      ybar interval=0.3,
      bar width=2pt,
      grid=major,
      xlabel={Empathy Scale Quality: Fantasy (FS)},
      ylabel={Total Score of Answers},
      ymin=0,
      ytick=data,
      xtick=data,
      axis x line=bottom,
      axis y line=left,
      enlarge x limits=0.1,
      symbolic x coords={awful,bad,average,good,excellent,ideal},
      xticklabel style={anchor=base,yshift=-0.5\baselineskip},
    ]
           \addplot[fill=yellow] coordinates {
               (awful,32) (bad,32)  (average,44)   (good,37)  (excellent,4) (ideal, 30)
  };
   \addplot[fill=white] coordinates {
      (awful,12)
       (bad,25)
        (average,20)
         (good,23)
        (excellent,31) (ideal, 30)      
      };

\legend{ Women, Men}
    \end{axis}
  \end{tikzpicture}

\subsection{Evidence of Spitefulness behavior}
In Table  \ref{ms} we compute the correlation between the subscale of empathy from the data collected from the participants. 
\begin{table}[htbp]
\begin{center}
\begin{tabular}{|c|c|c|c|c|} 
 \hline
Pearson correlation  & PT  & EC  & FS  & PD \\ \hline
PT & -& { \bf 0,81}& { \bf0,9382} & 0,2796  \\ \hline
EC & -&- & { \bf0,8709}& -0,3462  \\ \hline
FS & - &- & -& - \\ \hline
PD & -&- &- &-  \\ \hline
\end{tabular}
\end{center}
\caption{subscale correlation}
\label{ms}
\end{table}%

\begin{table}[htbp]
\begin{center}
\begin{tabular}{|c|c|} 
 \hline
 & Cooperation level \\ \hline
PT  + EC & 50\%  \\ \hline
PT + FS &  62,5\%   \\ \hline
PT + PD & 66,66 \% \\ \hline
EC + FS & 75\% \\ \hline
EC + PD & 50\% \\ \hline
\end{tabular}
\end{center}
\caption{Level of cooperation mixed scales}
\label{mscoop}
\end{table}%

The more refined result on the level of cooperation corresponding to the mixed IRI scale   is  given in Table  \ref{mscoop}. We can observe that a high level of cooperation associated to a high correlation coefficient correspond to  the Empathy-Altruism behavior  (namely  PT + FS  and EC + FS ).  A high level of cooperation associated to a negative or low correlation coefficient correspond to a sort of empathy-spitefulness behavior (namely PT + PD, EC + PD). In particular, the usage of empathy could be different across of the population and it is always in a positive sense.

The experiment reveals also that ``other" scale of empathy may be useful: (i) involvement of the users in technology is different across of the population \cite{involve}, and (ii) empathy anger may be correlated with the fact that some people are helping and some others have punishing desires \cite{anger}.  We leave these refined empathy concepts for future investigation.

\section{Conclusion}  \label{sec:conclusion}

Mean-field-type game theory is an emerging interdisciplinary toolbox with which one can describe situations where multiple persons make decisions and influence each other with state, type, actions and distributions of these. Psychological mean-field-type game theory is as an extension of those methods, with which one can design, analyze, identify how various psychological aspects, which classical models typically do not take into account, affect behaviour of the decision-makers. One can, for example, study the importance of empathy and emotions, for example  reciprocity, disappointment, regret, anger, shame, and guilt, that is the propensity to return favors, take revenge or being malicious or spiteful. Prior works on mean-field-type games have assumed that people's behavior is motivated solely by their own material payoff. Other aspects of motivation, for example the empathy and emotions, have been disregarded. But this is a major drawback  as empathy and emotions often influence behaviour and outcomes. 

In this paper, we have proposed and examined the role of empathy in mean-field-type games.  We established optimality system for such games when empathic player are involved. It is shown that empathic-altruism helps in reducing collision channel, securing the mean state and reducing electricity peak hours. Empathy-spitefulness of prosumers lowers electricity price and hence it helps consumers.  Empathy-altruism reduces inequality between the payoffs in mean-field-type games.  The experiment with 47 people carrying mobile devices   has demonstrated that using WiFi direct, D2D or other relaying technology on cell phones, tablets and laptops while moving or being in downtown or at airport degrades performance if the number of cooperators is not  sufficient enough, increasing the response time of the servers, particularly among the users who are far to the access points, and can lead to  interruption and call blocking. There is a need of coalition among a certain of number of nodes to maintain  a minimum connectivity level. The users' who are aware of such a situation may be empathetic. However, empathy can be used in different directions and different strategic ways: self-regarding, other-regarding, mutual-regarding, spitefulness, and indirect network effect etc.  The experiment reveals that more cooperation can be observed even in one-shot games when users' are empathetic, and this holds in both women population and men population.

Number of questions remain unanswered: (i) it would be interesting to examine the formation and the evolution of empathy as time goes, for example by means of learning process. (ii) time delayed empathy and forgiveness.  We have presented some of the extremes situations to illustrate clearly the influence of psychological factors. However, the interaction in engineering games  is not limited to empathy-altruism and empathy-spitefulness. There are multiple factors and multiple possibles cross-factors: one simple behavior to examine is the effect in the network when a user $i$ is helping user $j$ but not user $k$ and $j$ is helping $k$  but not $l$ etc. It is unclear who is helping whom in the multi-hop network through to the indirect path.
We leave these issues for  future investigation.

\appendix

\section*{Proof of Proposition \ref{prov1}:}
The evolution of the distribution of states  $m^s$ under state-and-mean-field feedback strategies is given by $$m_{t+1}^s(ds')= \int_s  q_{t+1}(ds' | \  s, m_t^s, m_t^a, a_t) m_t^s(ds)$$
This is a deterministic dynamics over $ \{0,1,\ldots, T\}.$
Since the expected payoff can be rewritten as a function of $m^s$ and the action, one can use a classical DPP with $m$ as a state.

Applying the classical dynamic programming principle (DPP) yields
$$\left\{ \begin{array}{c}
\hat{v}_{it}^{\lambda}(m_t^s)=\sup_{a'_i}\left\{ \hat{r}_{it}^{\lambda}( m_t^s,a'_{it},a_{-i,t}) \right.\\  \left. \quad \quad +\hat{v}_{i,t+1}^{\lambda}(m_{t+1}^s) \right\}\\
m_{t+1}^s(ds')= \int_s  q_{t+1}(ds' | \  s, m_t^s, m_t^a, a_t) m_t^s(ds).
\end{array}\right.
$$
This  completes the proof.

\section*{Proof of Proposition \ref{prov2}:}

We know from classical optimal control theory that  a pure optimal strategy may fail to exist in general.
However, one can extend  the  action space to the set of probability measures on $A$ and the underlying functions $\hat{r}_{it}^{\lambda}, \hat{g}_{iT}^{\lambda}, q_{t+1}$  can be extended. Then, the convexity of the action space is obtained.  In addition, if the continuity of the Hamiltonian holds  then one gets the existence of best response in behavioral (mixed) strategies. Using the multi-linearity property of the mixed extension procedure, one can use the Kakutani fixed-point theorem to obtain the existence of equilibria  in behavioral (mixed) strategies.

\section*{Proof of Proposition \ref{prov3}:}

We now show that if all the players are empathy-altruistic then the payoff gap is reduced across the entire network.  
Let $\lambda_{ij}=\lambda_{ji}=\lambda\in (0,1].$ Observing that \begin{eqnarray}
R_i^{\lambda}- R_j^{\lambda}&= & R_i-R_j+\sum_{k\neq i} \lambda_{ik} R_k-\sum_{k\neq j} \lambda_{jk} R_k\\
&=& R_i-R_j-\lambda (R_i-R_j)\\ &=&(1-\lambda) (R_i-R_j), \label{eq7}
\end{eqnarray}
Thus, from (\ref{eq7}) one obtains the following result:
If $R_i-R_j\neq 0,$ then the inequality ratio is
$$
\frac{| R_i^{\lambda}- R_j^{\lambda}|}{| R_i-R_j |}=1-\lambda < 1.
$$
which  completes the proof.

\begin{table*}
\centering
 \caption{IRI subscales. 
Extension of the empathy measure of Davis 1980,   Yarnold et al.1996  and Vitaglione et al. 2003. The star sign (*) denotes an opposite (reversed) counting/scoring.}
  \label{tab:table1}
  \begin{tabular}{|l|cccc|cccc|}
    \hline
    \multirow{1}{*}{Abridged item} &
      \multicolumn{4}{c}{Women (60\%)} &
      \multicolumn{4}{c}{Men(40\%)} \\
      & {PT} & EC& {FS} & {PD} &  {PT} & EC& {FS} & {PD}\\
      \hline
     (1) Daydream and fantasize (FS) &  &  &  &  & & & &  \\
    (2) Concerned with unfortunates (EC) &  & 0.6 &  &  & & & &  \\
    (3) Can't see  others' views$^{*}$ (PT) &  &  &  &  & & & &\\
     (4) Not sorry for others $^{*}$ (EC) &  &  & &  & & & &\\        
       (5)  Get involved in novels (FS)&  & & 0.8&  & & & &\\   
        (6)  Not-at-ease in emergencies (PD) & &  &  &  & & & & 0.7\\
         (7) Not caught-up in movies$^{*}$ (FS) &  & &  &  & & & &\\
          (8) Look at all sides in a fight (PT) &  & 0.9124  & 0.2444 &  & & & &\\
           (9) Feel protective of others  (EC) &  &  &  &  & &0.3 & &\\
           (10) Feel helpless when emotional (PD) &  &  &  &  & & & &\\
            (11) Imagine friend's perspective (PT) & &  &  &  & 0.8393& 0.824 & &\\
             (12) Don't get involved in books$^{*}$ (FS) & &  &  & & & & &\\
              (13) Remain calm if other's hurt $^{*}$ (PD)&  &  &  &  & & & &\\
               (14) Others' problems none mine$^{*}$ (EC) & &  &  & & & & &\\
                (15) If I'm right I won't argue$^{*}$ (PT)&  & &  &  & & & &\\
                 (16) Feel like movie character (FS) &  &  &  & & & & &\\
                  (17) Tense emotions scare me (PD) &  &  &  &  & & & &\\
                   (18) Don't feel pity for others $^{*}$ (EC)&  &  &  &  & & & &\\
                   (19) Effective in emergencies$^{*}$ (PD) &  &  &  &  & & & &\\
           (20) Touched by things I see  (EC)&  &  &  &  -0.3452& & & &\\
            (21) Two sides to every question  (PT)&  &  &  &  & & & &\\
             (22) Soft-hearted person (EC)&  &  &  &  & & & &\\
              (23) Feel like leading character (FS) & &  &  &  & & & &\\
               (44) Lose control in emergencies (PD)&  &  &  &  & & & &\\
                (25) Put myself in others' shoes (PT) &  &  &  &  & & & &\\
                 (26) Image novels were about me (FS)&  &  &  &  & & & &\\
                  (27) Other's problems destroy me (PD)& &  &  &  & & & &\\
                   (28) Put myself in other's place (PT)&  & 0.42&  &  & & & &\\
    \hline
  \end{tabular}
 \\  \vspace{1cm}
  \begin{tabular}{|l|cccc|cccc|}
    \hline
    \multirow{1}{*}{Decision outcome} &
      \multicolumn{4}{c}{Women (60\% of the whole population)} &
      \multicolumn{4}{c}{Men(40\%)} \\
      & {PT} & EC& {FS} & {PD} &  {PT} & EC& {FS} & {PD}\\
      \hline
     (FF)  &  10/28 & &  & & 7/19& & &  \\
    (FnF )  & &  &  &  & & & &  \\
     (nFF)  & &  &  &  & & & &  \\
    (nFnF)  & 3/28 &  & &  & 6/19& & &\\
     \hline
  \end{tabular}
\end{table*}
\newpage
\section*{Biography}

{\bf Giulia Rossi}  received  her Master degree  with summa cum laude  in  Clinical Psychology in 2009 from the University of Padova. She worked as independent researcher in the analysis and prevention of psychopathological diseases and in the intercultural expression of mental diseases.
Her research interests include behavioral game theory, social norms and the epistemic foundations of mean-field-type game theory. 	She is currently a research associate in the Learning \& Game Theory Laboratory at New York University Abu Dhabi.

{\bf Alain Tcheukam}  received  his PhD in   2013 in Computer Science and Engineering at the IMT Institute for Advanced Studies Lucca. 
His research interests include crowd flows, smart cities and mean-field-type optimization.
He received the  Federbim Valsecchi award 2015 for his contribution in design, modelling and analysis of smarter cities, and a best paper award 2016 from the International Conference on Electrical Energy and Networks. He is currently a postdoctoral researcher with Learning \& Game Theory Laboratory at New York University Abu Dhabi.

{\bf Hamidou Tembine} (S'06-M'10-SM'13) received the M.S. degree in Applied Mathematics from Ecole Polytechnique and the Ph.D. degree in Computer Science  from University of Avignon. His current research interests include evolutionary games, mean field stochastic games and applications.
 In 2014, Tembine received the IEEE ComSoc Outstanding Young Researcher Award for his promising research activities for the benefit of the society. He was the recipient of 7 best paper awards in the applications of game theory. Tembine is a prolific researcher and holds several scientific publications including magazines, letters, journals and conferences. He is author of the book on ``distributed strategic learning for engineers" (published by CRC Press, Taylor \& Francis 2012), and co-author of the book ``Game Theory and Learning in Wireless Networks" (Elsevier Academic Press). Tembine has been co-organizer of several scientific meetings on game theory in networking, wireless communications, transportation and smart energy systems.   He is a senior member of IEEE.


\begin{thebibliography}{99}
\bibitem{rmg} M. A. Khan, H. Tembine, Random matrix games in wireless networks,  IEEE Global High Tech Congress on Electronics (GHTCE 2012),
November 18-20, 2012, Shenzhen, China.
 \bibitem{t1}   G. Rossi, A. Tcheukam and H. Tembine,     
    How Much Does Users' Psychology Matter in Engineering Mean-Field-Type Games,
     Workshop on Game Theory and Experimental Methods
June 6-7, 2016,
Second University of Naples, Department of Economics Convento delle Dame Monache, Capua (Italy)



\bibitem{hannan} J. Hannan. Approximation to Bayes risk in repeated play. In M. Dresher, A. W. Tucker, and P. Wolfe (Eds.), Contributions to the Theory of Games, Vol. III, Ann. Math. Stud. 39, pp. 97-139. Princeton Univ. Press, 1957.
\bibitem{prest} Preston, S.D., \& de Waal, F.B.M. (2002). Empathy: its ultimate and proximate bases. Behavioral and Brain Sciences, 25(1), 1-71.

 \bibitem{psy1}Jan Grohn, Steffen Huck, Justin Mattias Valasek, A note on empathy in games, Journal of Economic Behavior \& Organization, Volume 108, December 2014, Pages 383-388,
\bibitem{psy2}Camerer CF (2003). Behavioral Game Theory. Princeton: Princeton University Press.
\bibitem{temftg}H. Tembine: Psychological mean-field-type games, Preprint, 2017.
\bibitem{temftg2}H. Tembine: Mean-field-type games, Preprint, 2017. 

\bibitem{psy3} Page K, Nowak M (2002), Empathy leads to fairness. Bull Math Biol 64: 1101-1116.
\bibitem{psy4}Pierpaolo Battigalli, Martin Dufwenberg, Dynamic psychological games, Journal of Economic Theory, vol. 144, Issue 1, January 2009, pp. 1-35.
\bibitem{psy5} G. Attanasi, R. Nagel, A survey of psychological games: Theoretical findings and experimental evidence, in: A. Innocenti, P. Sbriglia (Eds.), Games, Rationality and Behaviour. Essays on Behavioural Game Theory and Experiments, Palgrave McMillan, Houndmills, 2007, pp. 204-232.
\bibitem{psy6} J. Geanakoplos, D. Pearce, E. Stacchetti: Psychological games and sequential rationality, Games Econ. Behav. 1
(1989) 60-79.



\bibitem{jova82} Jovanovic, B. (1982): Selection and the Evolution of Industry, Econometrica 50, 649-670.

\bibitem{jova88}
{B. Jovanovic and R. W. Rosenthal} (1988). { Anonymous sequential games}, { Journal of Mathematical Economics}, 
vol. {17}, pp. {77-87}.


 \bibitem{marriage} D. Bauso, B. M. Dia, B. Djehiche, H. Tembine, R. Tempone (2014), Mean-Field Games for Marriage, PLoS One, 9(5): e94933. 

\bibitem{b1}  Andersson, D. and  Djehiche, B. (2010), A maximum principle for SDE's of mean-field type. \emph{Appl. Math. Optim. } 63(3), 341-356. 

\bibitem{book}H. Tembine: Distributed strategic learning for wireless engineers. CRC
Press/ Taylor \& Francis, 496 pages, 2012.

 \bibitem{ber1}Artinger F, Exadaktylos F, Koppel H, Saaksvuori L (2014), In others' shoes: do individual differences in empathy and theory of mind shape social preferences? PLoS One 9: e92844.
\bibitem{ber2} Karen M. Page , Martin A. Nowak: Empathy leads to fairness, Bulletin of Mathematical Biology, Nov. 2002, vol. 64, Issue 6, pp 1101-1116
 \bibitem{ber3}Pelligra, Vittorio, Empathy, guilt-aversion, and patterns of reciprocity. Journal of Neuroscience, Psychology, and Economics, vol 4(3), Aug 2011, 161-173.
 \bibitem{ber4}M Hartshorn, A Kaznatcheev, T Shultz, The evolutionary dominance of ethnocentric cooperation, Journal of Artificial Societies and Social Simulation 16 (3), 7, 2013.
 \bibitem{te1} H. Tembine:
Risk-sensitive mean-field-type games with Lp-norm drifts. Automatica, 59: 224-237 (2015)
  \bibitem{te2} H. Tembine:
Distributed massive MIMO network games: Risk and Altruism. CDC 2015: 3481-3486
\bibitem{te3} H. Tembine:
Nonasymptotic Mean-Field Games. IEEE Transactions on Cybernetics 44(12): 2744-2756 (2014).
\bibitem{nref1}
 Daeyeol Lee, Game theory and neural basis of social decision making, Nature Neuroscience 11, 404 - 409 (2008) 

\bibitem{nref2}  Alan G. Sanfey,  Social Decision-Making: Insights from Game Theory and Neuroscience,SCIENCE,  OCT 2007 : 598-602

\bibitem{nref3} Jean Decety, Empathy: A Game-Theoretic Approach, ISBN 9780262016612, MIT Press Ltd,  336 pages, 2011, 

\bibitem{nref4} K. Binmore, Playing Fair, volume 1 in  Game Theory and the Social Contract, 1994

\bibitem{nref5}Kenneth Binmore, Bargaining and fairness, PNAS, July  2014, vol. 111, 10785-10788
\bibitem{har} Harsanyi J.: Rational Behavior and Bargaining Equilibrium in Games and Social
Situations (Cambridge Univ Press, Cambridge, UK),  1977.
\bibitem{nref6}Trivers R.: The evolution of reciprocal altruism. Q Rev Biol 46:35-56, (1971).
\bibitem{nref7}Alan Kirman
and Miriam Teschl, 
Selfish or selfless? The role of empathy
in economics, Phil. Trans. R. Soc. B(2010), 365, 303-31

\bibitem{davis1}Davis, M. H. (1980). A multidimensional approach to individual differences in empathy. JSAS Catalog of Selected Documents in Psychology, 10, 85. 
\bibitem{temftg3}B. Djehiche, T. Basar, H. Tembine, Mean-Field-Type Game Theory, Springer, under preparation, 2017
\bibitem{temftg4} Alain Bensoussan, Boualem Djehiche, Hamidou Tembine, Phillip Yam, Risk-Sensitive Mean-Field-Type Control , Preprint, 2017, arXiv:1702.01369.

\bibitem{davis2} Davis, M. H. (1983). Measuring individual differences in empathy: Evidence for a multidimensional approach. Journal of Personality and Social Psychology, 44, 113- 126. 
\bibitem{batson01} C. Daniel Batson: Altruism in Humans, Oxford University Press, (2011).
\bibitem{refsmc1} Feng Zhou, Roger J. Jiao, Baiying Lei, Bilevel Game-Theoretic Optimization for Product Adoption Maximization Incorporating Social Network Effects, 
IEEE Transactions on Systems, Man, and Cybernetics: Systems 2016, pp. 1047- 1060 

\bibitem{refsmc2} Giuseppe Di Fatta , Guy Haworth: Skilloscopy: Bayesian Modeling of Decision Makers' Skill, 
IEEE Transactions on Systems, Man, and Cybernetics: Systems 2013, pp. 1290 - 1301 

\bibitem{refsmc3}  Sean B. Walker , Keith W. Hipel , Haiyan Xu: A Matrix Representation of Attitudes in Conflicts, 
IEEE Transactions on Systems, Man, and Cybernetics: Systems 2013, pp. 1328 - 1342 


\bibitem{anger}Guy D. Vitaglione and Mark A. Barnett: 
Assessing a New Dimension of Empathy: Empathic Anger as a Predictor of Helping and Punishing Desires, Motivation and Emotion, Vol. 27, No. 4, December 2003
\bibitem{involve} P. R. Yarnold , F. B. Bryant , S. D. Nightingale and G. J. Martin: Assessing physician empathy 
using the interpersonal reactivity index: A measurement model and cross-sectional analysis, Psychology, Health \& Medicine, 1:2, 207-221,  (1996).


\bibitem{rec1}Dufwenberg, M. and G. Kirchsteiger (2000) Reciprocity and wage undercutting, European Economic Review, 44: 1069-78.
 \bibitem{rec2}Dufwenberg, M. and G. Kirchsteiger (2004), A theory of sequential reciprocity, Games and Economic Behavior, 47: 268-98.


\end{thebibliography}
\end{document}